\documentclass[12pt,a4wide]{article}
\usepackage{amsmath,amssymb,amsfonts}

\newcommand{\eqa}{\begin{eqnarray}}
\newcommand{\ena}{\end{eqnarray}}
\newcommand{\topstar}[1]{\setlength{\unitlength}{1mm}
\begin{picture}(2,0)(-1,-1.4)
   \put(0,0){\makebox(0,0){$#1$}}
   \put(0,2.4){\makebox(0,0){\mbox{\tiny$\star$}}}
\end{picture}}
\begin{document}
\begin{center}
{\large {\bf Dark Matter Explanation from Quasi-Metric Gravity}}
\end{center}
\begin{center}
Dag {\O}stvang \\
{\em Department of Physics, Norwegian University of Science and Technology
(NTNU) \\
N-7491 Trondheim, Norway}
\end{center}
\begin{abstract}
The gravitational field of an isolated, axisymmetric flat disk of spinning 
dust is calculated approximatively in the weak-field limit of quasi-metric
gravity. Boundary conditions single out the exponential disk as a ``preferred''
physical surface density profile. Moreover, collective properties of the disk,
in the form of an extra ``associated induced'' surface density playing the role
of ``dark matter'' also emerge. Taken as idealized model of spiral galaxy thin 
disks, it is shown that including this ``dark matter'' into the model as a 
gravitating source, yields asymptotically flat rotation curves and a 
correspondence with MOND.
\\
\end{abstract}
\topmargin 0pt
\oddsidemargin 5mm
\renewcommand{\thefootnote}{\fnsymbol{footnote}}
\section{Introduction}
The concept of dark matter (DM), without which mainstream astrophysics and 
cosmology would not be observationally viable, is an acknowledged part of the 
modern scientific worldview. Yet all the observational evidence in favour of 
its existence is based on interpretations of astronomical data coming
from distance scales much larger than the solar system. In addition, despite 
its rather flexible nature, the DM proposal faces some real observational 
challenges. In particular galactic phenomenology, including spiral galaxy 
rotation curve shapes and the seemingly existence of a fixed acceleration
scale, seems difficult to understand in terms of DM.

This motivates the alternative approach of trying to explain anomalous galactic
observations from modified gravity. The most famous of these approaches is the
proposal known as MOND; i.e., a specific modification of Newtonian dynamics 
whenever the gravitational acceleration falls below a critical value $a_0$. 
Fact is that MOND does a very good job of modelling rotation curve shapes of 
spiral galaxies just from their visible matter content, using only one freely 
variable parameter $a_0$. That this feat is possible at all, makes the DM 
approach look suspicious in comparison since no {\em a priori} reason is given 
for why a much greater variety of rotation curve shapes is not observed. 
However, MOND does not work so well on larger distance scales such as clusters 
of galaxies. Besides, relativistic extensions of MOND involve arbitrary extra 
fields reducing their predictive powers and thus their advantages vis-a-vis DM.

Rather than trying to modify gravity by introducing extra fields tailored to 
fit galactic phenomenology, a less contrived approach would be to adopt a 
general alternative framework of relativistic gravity and see if galactic
phenomena can be correctly predicted from first principles. Such an approach
can hardly be made to work within the traditional framework of metric theories
of gravity [1]. However, another possibility is the so-called quasi-metric
framework (QMF) published some time ago [2, 3]. Quasi-metric gravity is rather 
arcane and it has not yet been shown to be viable. On the other hand, nor is it
in obvious conflict with observations, even if it would seem so at first 
glance. The QMF is also radical inasmuch as the role of the cosmic expansion is
described very differently from its counterpart in metric theories of gravity. 
That is, the QMF describes the cosmic expansion as a general phenomenon not 
depending on space-time's causal structure. This yields a number of (unique)
predictions.

Specifically, the QMF predicts quantitatively, and from first principles, how 
the cosmic expansion influences gravitationally bound systems. In particular, 
the QMF predicts how the cosmic expansion should affect the solar system and 
that these effects should be observable. A number of observed solar system 
phenomena have been reinterpreted and shown to be in good agreement with these 
predictions as long as no extra theory-dependent assumptions are made [4]. 
Unfortunately, these results indicate that general relativity (GR) is 
fundamentally flawed and that interpretations of some solar system observations
are crucially theory-dependent. Well-known results based on traditional 
interpretations of such observations may therefore be unreliable. Moreover,
questioning traditional interpretations is probably why quasi-metric gravity is
ignored by the scientific community, despite making a number of successful 
predictions not shared with GR [4]. Besides, if it turns out that quasi-metric
gravity is able to explain galactic phenomena from first principles as well, 
this will contribute to undercutting the relevance of the DM concept. In that 
case the legitimacy of mainstream astrophysics will be weakened even further. 

The goal of showing the compatibility of the QMF with galactic phenomena, 
without the assumption of DM, will be partially fulfilled in this paper. 
That is, we calculate the gravitational field of a very thin, rotating 
disk of dust in the non-relativistic limit of quasi-metric gravity. Taking the 
spinning disk as an idealized model of a spiral galaxy, we indicate a solution 
of some galactic observations without the need of DM. The correspondence of 
this solution to that of MOND is also discussed. But first, in the next 
section, a brief introduction to quasi-metric gravity will be presented.
\section{Basic quasi-metric gravity}
The QMF has been described in detail elsewhere [2-4]. Here we include only
a minimum of motivation and general formulae necessary to do the
calculations presented in later sections.

The basic motivation for introducing the QMF is of a very general philosophical
nature. That is, traditional field theories consist of two independent parts;
field equations and initial conditions. This general form ensures that field 
theories can in principle be applied to all physical systems within their
domain of validity. But for cosmology this general flexibility is a liability; 
since the Universe is unique from an observational point of view, it is in 
principle impossible to have observational knowledge of alternatives to cosmic
initial conditions, global evolution and structure. Any diversity of such
possibilities represents a serious limitation to what can be known in principle,
and should be avoided if possible. That is, since the Universe is 
observationally unique, so should the nature of its global evolution be.

It turns out that, to construct a general mathematical framework fulfilling
this requirement, one is pretty much led to the geometrical structure of the
QMF. Moreover, the QMF accomplishes this requirement by describing the global
cosmic expansion as an absolute, prior-geometric phenomenon not being part of
space-time's causal structure. In this way, the cosmic expansion does not 
depend on gravitational field equations or initial conditions, meaning that the
Universe is not described as a purely dynamical system. That is, the QMF 
describes the cosmic expansion as ``new physics'' with no correspondence (in 
any limit) to the standard Lorentzian space-time framework.

Similarly to the Robertson-Walker (RW) manifolds in GR, the cosmic expansion in
the QMF is defined by means of a family of ``preferred'' observers, the 
so-called {\em fundamental observers} (FOs). A further similarity with the 
RW-manifolds is the existence of a {\em global time function $t$}, such that 
$t$ splits up space-time into a ``distinguished'' set of spatial hypersurfaces,
the so-called {\em fundamental hypersurfaces} (FHSs). But since the cosmic 
expansion in the QMF by hypothesis is not part of space-time's causal 
structure, $t$ cannot be an ordinary time coordinate on a Lorentzian manifold.
Rather, $t$ should play the role of an independent evolution parameter 
parametrizing any change in the space-time geometry that has to do with the 
cosmic expansion. On the other hand, space-time must also be equipped with a 
causal structure in the form of a Lorentzian manifold. This Lorentzian manifold
must also accommodate the FOs and the FHSs, which means that its topology 
should allow the existence of a global ordinary time coordinate $x^0$. (Note 
that, to ensure the uniqueness of this construction, the FHSs must be compact.)

Taking into account the above considerations, the geometrical basis of the QMF
can now be defined. That is, the basic geometrical structure underlying the QMF
consists of a 5-dimensional differentiable manifold with topology 
${\cal M}{\times}{\bf R}_1$, where ${\cal M}={\cal S}{\times}{\bf R}_2$ is a 
Lorentzian space-time manifold, ${\bf R}_1$ and ${\bf R}_2$ both denote the 
real line and ${\cal S}$ is a compact 3-dimensional manifold (without 
boundaries). This means that, in addition to the usual time dimension and 3 
space dimensions, there is an extra time dimension represented by {\em the 
global time function} $t$. Moreover, the manifold ${\cal M}{\times}{\bf R}_1$ 
is equipped with two degenerate 5-dimensional (covariant) metrics 
${\bf {\bar g}}_t$ and ${\bf g}_t$, where the degeneracies are determined by 
the conditions ${\bf {\bar g}}_t({\frac{\partial}{{\partial}t}},{\cdot})
{\equiv}0$ and ${\bf g}_t({\frac{\partial}{{\partial}t}},{\cdot}){\equiv}0$, 
respectively. The metric ${\bf {\bar g}}_t$ is directly coupled to matter 
fields via gravitational field equations, whereas the ``physical'' metric 
${\bf g}_t$ can be constructed from ${\bf {\bar g}}_t$ in a way described in 
refs. [2, 3]. (See also section 6.) Note that ${\bf {\bar g}}_t$ and 
${\bf g}_t$ have the property that the FOs always move orthogonally to the 
FHSs.

To reduce space-time to 4 dimensions, one obtains the quasi-metric space-time
manifold $\cal N$ by slicing the submanifold determined by the equation 
$x^0=ct$ out of the 5-dimensional differentiable manifold ${\cal M}{\times}
{\bf R}_1$. It is essential that this slicing is unique since the two global
time coordinates should be physically equivalent; the only
reason to separate between them is that they are designed to parameterize 
fundamentally different physical phenomena. Since the geometric structure on
$\cal N$ is inherited from that on ${\cal M}{\times}{\bf R}_1$ just by 
restricting the fields to $\cal N$ (no projections), the 5-dimensional 
degenerate metric fields ${\bf {\bar g}}_t$ and ${\bf g}_t$ may be regarded as 
one-parameter families of Lorentzian 4-metrics on $\cal N$ (this terminology is
merely a matter of semantics). Furthermore, there exists a set of particular 
coordinate systems especially well adapted to the geometrical structure of 
quasi-metric space-time, {\em the global time coordinate systems (GTCSs)}. 
A coordinate system is a GTCS iff the time coordinate $x^0$ is related to $t$ 
via $x^0=ct$ in ${\cal N}$.

Note that the existence of a ``preferred frame'' and a unique class of FOs are 
intrinsic, {\em global} geometric properties of quasi-metric space-time. In 
practice, this means that the FOs should be identified with observers being at 
rest, on average, with respect to the cosmic rest frame. However, for an 
isolated system that is small compared to cosmological scales and is moving 
slowly with respect to the cosmic rest frame, one may ignore the global 
curvature of space and pretend that it is asymptotically flat. Then the rest 
frame of the system may be used as an approximate substitute for the cosmic 
rest frame [2, 3]. Moreover, an alternative class of observers being at rest, 
on average, with respect to the isolated system's rest frame, may be used as 
an approximate substitute for the class of FOs. Any ``preferred frame''-effects
missed by said approximation should be at most of post-Newtonian order.

Since the main role of $t$ is to describe how the cosmic expansion influences 
space-time geometry, $t$ should enter ${\bf {\bar g}}_t$ and ${\bf g}_t$ 
explicitly as a scale factor (in addition to any implicit dependence on $t$). 
However, unlike its counterpart in the RW-models, this scale factor cannot be 
calculated from gravitational field equations, but must be an ``absolute'' 
quantity. Since the form of the scale factor should not introduce any extra 
arbitrary scale or parameter, the only possible option for a scale factor with 
the dimension of length is to set it equal to $ct$. This scale factor may be 
multiplied with a second, dimensionless scale factor taking into account 
the most important effects of gravity for weak gravitational fields. But since 
the weak-field geometry of the FHSs in $({\cal N},{\bf {\bar g}}_t)$ is 
postulated to represent a measure of gravitational scales in terms of atomic 
units [2, 3], any extra dimensionless scale factor should enter 
${\bf {\bar g}}_t$ as a conformal factor.

Furthermore, since there is no reason to introduce any nontrivial spatial 
topology, the global basic geometry of the FHSs (neglecting the effects of 
gravity) should be that of the 3-sphere ${\bf S}^3$. This introduces some
prior 3-geometry of the FHSs (see equation (8) below). It then turns out that 
the most general form of ${\bf {\bar g}}_t$ (expressed in a GTCS) can be 
written as a family of line elements (using Einstein's summation convention)
\eqa
{\overline {ds}}_t^2={\bar N}_t^2{\Big \{ }
[{\bar N}_{(t)}^k{\bar N}_{(t)}^s{\tilde h}_{(t)ks}-1](dx^0)^2+
2{\frac{t}{t_0}}{\bar N}_{(t)}^k{\tilde h}_{(t)ks}dx^sdx^0+
{\frac{t^2}{t_0^2}}{\tilde h}_{(t)ks}dx^kdx^s{\Big \} },
\ena
where $t_0$ is some arbitrary reference epoch setting the scale of the spatial 
coordinates, ${\bar N}_t$ is the family of lapse functions of the FOs and 
${\frac{t_0}{t}}{\bar N}^k_{(t)}$ are the components of the shift vector family 
of the FOs in $({\cal N},{\bf {\bar g}}_t)$. Moreover, $d{\bar{\sigma}}_t^2
{\equiv}{\bar h}_{(t)ks}dx^kdx^s{\equiv}
{\frac{t^2}{t_0^2}}{\bar N}_t^2{\tilde h}_{(t)ks}dx^kdx^s$ is the line element 
family corresponding to the spatial metric family ${\bf {\bar h}}_t$ intrinsic 
to the FHSs.

The affine structure constructed on ${\cal M}{\times}{\bf R}_1$ limits any 
possible $t$-dependence of the quantities present in equation (1) (see below).
Specifically, there are restricions on how ${\bar N}^i_{(t)}$ may depend on $t$ 
given the quantities ${\tilde h}_{(t)ks},_t$ [2, 3] (here, a comma denotes 
taking a partial derivative). Also note that, since the 5-dimensional metrics 
are degenerate, there are no components of lapse and shift in the $t$-direction 
(i.e., there is no motion and proper time does not elapse along the 
$t$-direction).

Next, $({\cal N},{\bf {\bar g}}_t)$ and $({\cal N},{\bf g}_t)$ are equipped
with linear and symmetric connections ${\topstar {\bar {\nabla}}}$ and
${\topstar {\nabla}}$, respectively. These connections are identified with the
usual Levi-Civita connection for constant $t$, yielding the standard connection
coefficients, while the rest of the connection coefficients are determined by
the requirements
\eqa
{\topstar {\bar {\nabla}}}_{\frac{\partial}{{\partial}t}}{\bf {\bar g}}_t=0,
\qquad
{\topstar {\bar {\nabla}}}_{\frac{\partial}{{\partial}t}}{\bf {\bar n}}_t=0,
\qquad
{\topstar {\nabla}}_{\frac{\partial}{{\partial}t}}{\bf g}_t=0, \qquad
{\topstar {\nabla}}_{\frac{\partial}{{\partial}t}}{\bf n}_t=0,
\ena
where ${\bf {\bar n}}_t$ and ${\bf n}_t$ are families of unit normal vector 
fields in $({\cal N},{\bf {\bar g}}_t)$ and $({\cal N},{\bf g}_t)$,
respectively. The requirements shown in equation (2) yield some extra, 
potentially nonzero connection coefficients; see refs. [2, 3] for explicit
expressions.

Now it turns out that a full coupling of {\em the active stress-energy tensor} 
${\bf T}_t$ to space-time curvature cannot exist [2, 3]. Besides, only that 
part of the space-time curvature obtained by holding $t$ constant should couple
to matter. Even more radical is the requirement from the QMF that the 
gravitational coupling of matter to space-time geometry must be nonuniversal, 
i.e., that there must exist two different gravitational coupling parameters 
$G_t^{\rm B}$ and $G_t^{\rm S}$ and that these must be variable [2, 3]. That is, 
$G_t^{\rm B}$ and $G_t^{\rm S}$ couple, respectively, the active electromagnetic 
stress-energy tensor ${\bf T}^{\rm (EM)}_t$ and the active stress-energy tensor 
for material sources ${\bf T}^{\rm (MA)}_t$, to space-time geometry.

It turns out that a set of field equations, somewhat similar to a subset of the 
projected Einstein field equations in canonical GR, can be tailored to 
${\bf {\bar g}}_t$, so that {\em partial} gravitational couplings of 
${\bf T}^{\rm (EM)}_t$ and ${\bf T}^{\rm (MA)}_t$ to space-time geometry exist 
[2, 3]. These field equations read (expressed in a GTCS)
\eqa
2{\bar R}_{(t){\bar {\perp}}{\bar {\perp}}}&=&
2(c^{-2}{\bar a}_{{\cal F}{\mid}i}^i+
c^{-4}{\bar a}_{{\cal F}i}{\bar a}_{\cal F}^i-
{\bar K}_{(t)ik}{\bar K}_{(t)}^{ik}+
{\cal L}_{{\bf {\bar n}}_t}{\bar K}_t) \nonumber \\
&=&{\kappa}^{\rm B}(T^{\rm (EM)}_{(t){\bar {\perp}}{\bar {\perp}}}
+{\hat T}_{(t)i}^{{\rm (EM)}i})+
{\kappa}^{\rm S}(T^{\rm (MA)}_{(t){\bar {\perp}}{\bar {\perp}}}
+{\hat T}_{(t)i}^{{\rm (MA)}i}),
\qquad c^{-2}{\bar a}_{{\cal F}j}{\equiv}{\frac{{\bar N}_t,_j}{{\bar N}_t}},
\ena
\eqa
{\bar R}_{(t)j{\bar {\perp}}}+{\Big (}{\frac{{\bar h}_{(t)}^{ik}}{{\bar N}_t}}
{\frac{\partial}{{\partial}x^0}}{\bar h}_{(t)ij}{\Big )}_{{\mid}k}-
{\Big (}{\frac{{\bar h}_{(t)}^{ik}}{{\bar N}_t}}
{\frac{\partial}{{\partial}x^0}}{\bar h}_{(t)ik}{\Big )},_j
={\kappa}^{\rm B}T^{{\rm (EM)}}_{(t)j{\bar {\perp}}}
+{\kappa}^{\rm S}T^{\rm (MA)}_{(t)j{\bar {\perp}}}.
\ena
Here, ${\bf {\bar R}}_t$ is the Ricci tensor family corresponding to the metric
family ${\bf {\bar g}}_t$ and the symbol '${\bar {\perp}}$' denotes a scalar
product with $-{\bf {\bar n}}_t$. Moreover, ${\cal L}_{{\bf {\bar n}}_t}$ denotes 
a projected Lie derivative in the direction normal to the FHSs, 
${\bf {\bar K}}_t$ denotes the extrinsic curvature tensor family (with trace 
${\bar K}_t$) of the FHSs, a ``hat'' denotes an object projected into the FHSs 
and the symbol '${\mid}$' denotes taking a spatial covariant derivative. 
Finally, ${\kappa}^{\rm B}{\equiv}8{\pi}G^{\rm B}/c^4$ and
${\kappa}^{\rm S}{\equiv}8{\pi}G^{\rm S}/c^4$,  where $G^{\rm B}$ and $G^{\rm S}$ are 
by convention chosen as the values of $G^{\rm B}_t$ and $G^{\rm S}_t$, 
respectively, measured in (hypothetical) local gravitational experiments in an 
empty universe at epoch $t_0$. In addition to equations (3) and (4), we also 
have the extra field equation (without any independent coupling to ${\bf T}_t$)
[2, 3]
\eqa
{\frac{1}{{\bar N}_t}}{\cal L}_{{\bar N}_t{\bf {\bar n}}_t}{\bar K}_{(t)ij}
+{\bar K}_t{\bar K}_{(t)ij}-{\tilde H}_{(t)ij} \nonumber \\
={\frac{1}{3}}{\Big [}{\bar R}_{(t){\bar {\perp}}{\bar {\perp}}}
-{\bar K}_{(t)ks}{\bar K}_{(t)}^{ks}+{\bar K}_t^2
-c^{-2}{\bar a}_{{\cal F}{\mid}s}^s
-c^{-4}{\bar a}_{{\cal F}}^s{\bar a}_{{\cal F}s}+{\frac{3}{(ct{\bar N}_t)^2}}
{\Big ]}{\bar h}_{(t)ij},
\ena
where ${\tilde H}_{(t)ij}$ are the components of the Einstein tensor family
${\bf {\tilde H}}_t$ obtained from the metric family ${\bf {\tilde h}}_t$ (note
that ${\bf {\bar h}}_t{\equiv}{\frac{t^2}{t_0^2}}
{\bar N}_t^2{\bf {\tilde h}}_t$). For an isolated system and sufficiently weak 
gravitational fields, equation (5) contributes only to higher order terms in 
small quantities. Therefore, for such cases, it may often be a good 
approximation to ignore equation (5) altogether.

We now list some formulae being useful when trying to solve the field 
equations. First, a well-known (except for the $t$-dependence) 
formula for ${\bf {\bar K}}_t$ coming from canonical GR is
\eqa
{\bar K}_{(t)ij}={\frac{1}{2{\bar N}_t}}{\Big [}{\frac{t}{t_0}}
({\bar N}_{(t)i{\mid}j}+{\bar N}_{(t)j{\mid}i})-
{\frac{\partial}{{\partial}x^0}}{\bar h}_{(t)ij}{\Big ]}.
\ena 
Note that in general, we have that
${\bar R}_{(t)j{\bar {\perp}}}={\bar K}_{(t)j{\mid}i}^i-{\bar K}_t,_j$.
Other useful formulae consistent with equation (1) are
\eqa
{\bar H}_{(t)ij}=c^{-2}{\Big (}{\bar a}_{{\cal F}{\mid}k}^k- 
{\frac{1}{{\bar N}_t^2t^2}}{\Big )}{\bar h}_{(t)ij}-c^{-4}
{\bar a}_{{\cal F}i}{\bar a}_{{\cal F}j}-c^{-2}
{\bar a}_{{\cal F}i{\mid}j}+{\tilde H}_{(t)ij},
\ena
\eqa
{\bar P}_t={\frac{6}{({\bar N}_tct)^2}}+2c^{-4}{\bar a}_{{\cal F}k}
{\bar a}_{\cal F}^k-4c^{-2}{\bar a}_{{\cal F}{\mid}k}^k, \qquad
{\tilde P}_t={\frac{6}{(ct_0)^2}},
\ena
where ${\bf {\bar H}}_t$ is the Einstein tensor family intrinsic to the FHSs
in $({\cal N},{\bf {\bar g}}_t)$ and ${\bar P}_t$ is the corresponding 
curvature scalar family. Similarly, ${\tilde P}_t$ is the curvature scalar
associated with the metric family ${\bf {\tilde h}}_t$. Note that 
${\tilde P}_t$ represents prior 3-geometry of the FHSs and that it does not 
depend explicitly on $t$. Several more useful formulae relevant for 
quasi-metric gravity may be found in [2, 3].

As should be clear by now, the most characteristic feature of the QMF is the
existence of a nonmetric sector of quasi-metric space-time, describing the
cosmic expansion as a general physical phenomenon not depending on space-time's
causal structure. This feature makes the QMF unique and distinguishes it from
all other, more ``standard'' theories of modified gravity where it is assumed
that space-time is still modelled as a Lorentzian manifold, but with field
equations derived from some alternative action. But even within its metric 
sector, the QMF differs from other modified gravity theories by the existence 
of a (nonpropagating) {\em indirectly coupled dynamical degree of freedom} 
represented by the transformation ${\bf {\bar g}}_t{\rightarrow}{\bf g}_t$ 
[2, 3]. This transformation is of a purely geometrical nature and its effects 
are at the post-Newtonian level, so it does not have much relevance for the 
results presented in this paper (but see section 6).

At this point, it should be clear that the basic structure of the QMF has 
nothing whatsoever in common with that of MOND or any of its relativistic
extensions, and in particular not in the weak-field approximation. How then is
it possible to get a correspondence with MOND at all? The answer to this 
question is that such a correspondence could occur for special cases, even if 
no general weak-field correspondence exists. Indeed, it is shown in this paper 
that such a correspondence occurs as a nonlocal, {\em collective} phenomenon 
for the case of a flat disk. And as we shall see, the combination of a flat 
matter distribution and a ``small'' Universe with spherical spatial geometry 
at cosmic scales is what makes this correspondence possible.
\section{Axisymmetric, metrically stationary, flat systems}
\subsection{Real weak-field approximate solution}
It was found in [5] that no solutions of the field equations (3)-(5) exist for
an isolated, axially symmetric, metrically stationary, rotating system. 
However, an approximate exterior solution of equations (3) and (4) can be found
for weak gravitational fields (ignoring equation (5)). For this case, 
introducing a spherically symmetric GTCS 
${\{ }x^0,{\rho},{\theta},{\phi}{\} }$, where ${\rho}$ is an isotropic radial 
coordinate and ${\bar N}_t$ and ${\bar N}_{(t){\phi}}$ do not depend on ${\phi}$,
equation (1) takes the approximate form [5]
\eqa
{\overline{ds}}^2_t={\bar B}{\Big [}-(1-{\bar V}^2{\rho}^2{\sin}^2{\theta})
(dx^0)^2+2{\frac{t}{t_0}}{\bar V}{\rho}^2{\sin}^2{\theta}d{\phi}dx^0 
+{\frac{t^2}{t_0^2}}{\Big (}{\frac{d{\rho}^2}{1-{\frac{{\rho}^2}
{{\Xi}_0^2}}}}+{\rho}^2d{\Omega}^2{\Big )}{\Big ]},
\ena
where $d{\Omega}^2{\equiv}d{\theta}^2+{\sin}^2{\theta}d{\phi}^2$, 
${\Xi}_0{\equiv}ct_0$, ${\bar B}{\equiv}{\bar N}_t^2$ and ${\bar V}{\equiv}
{\frac{{\bar N}_{(t){\phi}}}{{\bar B}{\rho}^2{\sin}^2{\theta}}}$. The field
equations (3), (4) exterior to the source then read [5]
\eqa
(1-{\frac{{\rho}^2}{{\Xi}_0^2}}){\bar B},_{{\rho}{\rho}}+{\frac{1}{{\rho}^2}}
{\bar B},_{{\theta}{\theta}}&+&{\frac{2}{\rho}}
(1-{\frac{3{\rho}^2}{2{\Xi}_0^2}})
{\bar B},_{\rho}+{\frac{{\cot}{\theta}}{{\rho}^2}}{\bar B},_{\theta}
\nonumber \\
&=&{\bar B}{\rho}^2{\sin}^2{\theta}{\Big [}(1-{\frac{{\rho}^2}{{\Xi}_0^2}})
({\bar V},_{\rho})^2+{\frac{1}{{\rho}^2}}({\bar V},_{\theta})^2{\Big ]},
\ena
\eqa
(1-{\frac{{\rho}^2}{{\Xi}_0^2}}){\bar V},_{{\rho}{\rho}}+{\frac{1}{{\rho}^2}}
{\bar V},_{{\theta}{\theta}}+{\Big [}{\frac{4}{\rho}}-
{\frac{5{\rho}}{{\Xi}_0^2}}+(1-{\frac{{\rho}^2}{{\Xi}_0^2}})
{\frac{{\bar B},_{\rho}}{\bar B}}{\Big ]}{\bar V},_{\rho} 
+{\Big [}3{\cot}{\theta}+{\frac{{\bar B},_{\theta}}{\bar B}}{\Big ]}
{\frac{1}{{\rho}^2}}{\bar V},_{\theta}=0.
\ena
For flat systems, it is convenient to switch to a cylindrical GTCS
$(x^0,{\xi},z,{\phi})$, where ${\xi}{\equiv}{\rho}{\sin}{\theta}$,
$z{\equiv}{\rho}{\cos}{\theta}$. Then equation (9) becomes
\eqa
{\overline{ds}}^2_t={\bar B}{\Big [}-(1-{\bar V}^2{\xi}^2)
(dx^0)^2+2{\frac{t}{t_0}}{\bar V}{\xi}^2d{\phi}dx^0 \nonumber \\ 
+{\frac{t^2}{t_0^2}}{\Big (}
(1-{\frac{{\xi}^2+z^2}{{\Xi}_0^2}})^{-1}{\Big [}(1-{\frac{z^2}{{\Xi}_0^2}})
d{\xi}^2+(1-{\frac{{\xi}^2}{{\Xi}_0^2}})dz^2+
2{\frac{{\xi}z}{{\Xi}_0^2}}d{\xi}dz{\Big ]}+{\xi}^2d{\phi}^2{\Big )}{\Big ]},
\ena
and the field equations (10), (11) read
\eqa
(1-{\frac{{\xi}^2}{{\Xi}_0^2}}){\bar B},_{{\xi}{\xi}}&+&
(1-{\frac{z^2}{{\Xi}_0^2}}){\bar B},_{zz}-{\frac{2{\xi}z}{{\Xi}_0^2}}
{\bar B},_{{\xi}z}+{\frac{1}{\xi}}(1-{\frac{3{\xi}^2}{{\Xi}_0^2}})
{\bar B},_{\xi}-{\frac{3z}{{\Xi}_0^2}}{\bar B},_z
\nonumber \\
&=&{\bar B}{\xi}^2{\Big [}(1-{\frac{{\xi}^2}{{\Xi}_0^2}})
({\bar V},_{\xi})^2+(1-{\frac{z^2}{{\Xi}_0^2}})({\bar V},_z)^2
-{\frac{2{\xi}z}{{\Xi}_0^2}}{\bar V},_{\xi}{\bar V},_z{\Big ]},
\ena
\eqa
(1-{\frac{{\xi}^2}{{\Xi}_0^2}}){\bar V},_{{\xi}{\xi}}&+&
(1-{\frac{z^2}{{\Xi}_0^2}}){\bar V},_{zz}-
{\frac{2{\xi}z}{{\Xi}_0^2}}{\bar V},_{{\xi}z}
+{\Big [}(1-{\frac{{\xi}^2}{{\Xi}_0^2}}){\frac{{\bar B},_{\xi}}{\bar B}}
-{\frac{{\xi}z}{{\Xi}_0^2}}{\frac{{\bar B},_z}{\bar B}} \nonumber \\
&&+{\frac{1}{\xi}}(3-{\frac{5{\xi}^2}{{\Xi}_0^2}}){\Big ]}{\bar V},_{\xi}
+{\Big [}(1-{\frac{z^2}{{\Xi}_0^2}}){\frac{{\bar B},_z}{\bar B}}
-{\frac{{\xi}z}{{\Xi}_0^2}}{\frac{{\bar B},_{\xi}}{\bar B}}
-{\frac{5z}{{\Xi}_0^2}}{\Big ]}{\bar V},_z=0.
\ena
We now assume that the gravitational field is so weak that we may ignore the 
system's rotation and thus set ${\bar V}=0$. This implies that equation (14) 
becomes vacuous and that the right hand side of equation (13) vanishes. To 
find ${\bar B}({\xi},z)$, we are thus left to solve equation (13) with the 
right hand side equal to zero. Note that this is equivalent to solving the 
Laplace equation on a subset of the 3-sphere. We also assume that the 
gravitational source contains a negligible amount of electromagnetic field 
energy so that the gravitational coupling is adequately described using the 
gravitational coupling parameter $G^{\rm S}_t$.

Unfortunately, equation (13) is a non-separable partial differential equation
(PDE). This means that there is not much hope of finding exact solutions. 
However, one may try to find approximate series solutions for $|z|<<{\Xi}_0$. 
The series expansions in terms of $|z|$ should then take the same form as for 
the corresponding Newtonian problem, recovered by letting ${\Xi}_0{\rightarrow}
{\infty}$ in equation (13). For that case, one gets a separable PDE in the 
Newtonian potential. That problem was solved many years ago [6, 7], yielding a 
continuous spectrum of solutions ${\Phi}_k({\xi},z){\ }{\propto}{\ }
J_0(k{\xi}){\exp}(-k|z|)$, where $J_0(k{\xi})$ is a Bessel function of the 
first kind. We are thus led, via correspondence with the Newtonian case, to 
try solutions ${\bar B}({\xi},z)=1+{\frac{2}{c^2}}{\Phi}({\xi},z)$
built from mode solutions of the form
\eqa
{\Phi}_k({\xi},z)={\Phi}_k({\xi}){\Big (}1-k|z|+
{\frac{1}{2!}}{\alpha}_k(kz)^2+{\cdots}{\Big )},
\ena
where the ${\alpha}_k$ are constants (but higher order coefficients will in 
general depend on $\xi$). The reason why the ${\alpha}_k$ do not depend on 
$\xi$, is that any deviation from Euclidean space for small $|z|$ occurs at 
order $|z|^3$ and higher. That is, by integrating the spatial line element a 
distance $|z|<<{\Xi}_0$ in the $z$-direction we find (for ${\bar B}{\approx}1$)
\eqa
{\sqrt{1-{\frac{{\xi}^2}{{\Xi}_0^2}}}}{\int}^z_{{\!}{\!}{\!}0}
{\frac{dz'}{\sqrt{1-{\frac{{\xi}^2+{z'}^2}{{\Xi}_0^2}}}}}=
z{\Big (}1+{\frac{1}{6(1-{\frac{{\xi}^2}{{\Xi}_0^2}})}}
{\frac{z^2}{{\Xi}_0^2}}+{\cdots}{\Big )}.
\ena
We now insert equation (15) into equation (13) (with the right hand side set to 
zero) and collect terms to get separate equations for each power of $|z|$. 
To lowest order, the terms independent of $|z|$ yield an equation for 
${\Phi}_k({\xi})$, i.e.
\eqa
(1-{\frac{{\xi}^2}{{\Xi}_0^2}}){{\Phi}_k,}_{{\xi}{\xi}}+
{\frac{1}{\xi}}(1-{\frac{3{\xi}^2}{{\Xi}_0^2}}){{\Phi}_k},_{\xi}+
{\alpha}_kk^2{\Phi}_k=0.
\ena
Similarly, collecting terms of first order in $|z|$ yields an equation 
determining the third order coefficient of the series expansion in equation 
(15), and so on for each order in $|z|$.

Now, since $k$ is interpreted as a wavenumber on the 3-sphere, it must have a
minimum value $k_0={\frac{1}{{\Xi}_0}}$ corresponding to the maximum
value ${\xi}_{\rm{max}}={\Xi}_0$. This indicates that, rather than the 
continuous spectrum of solutions found for the Newtonian case, the solution of 
equation (17) should involve a {\em discrete} spectrum of solutions
${\Phi}_n({\xi}){\equiv}{\Phi}_{k_n}({\xi})$. In fact the general solution of 
equation (17) is (with ${\alpha}_n{\equiv}{\alpha}_{k_n}$)
\eqa
c^{-2}{\Phi}_n({\xi})=-C_nP_n(u)-C^{\rm i}_nQ_n(u), \qquad 
n=({\sqrt{{\alpha}_nk_n^2{\Xi}_0^2+1}}-1)/2, \qquad
u{\equiv}1-2{\frac{{\xi}^2}{{\Xi}_0^2}},
\ena
where $C_n$, $C^{\rm i}_n$ are (dimensionless) constants and $P_n(u)$, $Q_n(u)$ 
are Legendre functions of the first and second kind, respectively. (Note
that the mode solutions $C^{\rm i}_nQ_n(u)$ diverge logarithmically when 
$u{\rightarrow}{\pm}1$.) From equation (18) we see that we get a discrete 
spectrum of solutions if $n$ is required to be a non-negative integer. 
Moreover, we require that $k_n$ should be an integer multiple of $k_0$, and 
that in the continuum limit $ \lim_{n{\rightarrow}{\infty}}{\alpha}_n=1$, so we 
must choose
\eqa
{\alpha}_n=1-{\frac{1}{k_n^2{\Xi}_0^2}}, \quad \Rightarrow \quad
k_n={\frac{2n+1}{{\Xi}_0}}, \quad n=0,1,2,{\dots},
\ena
and this choice will also be consistent with the indicated value of $k_0$. Note
that, to get the correspondence with the Newtonian case, one may choose
$C^{\rm i}_n=0$ and $C_n=e^{-s}{\frac{s^n}{n!}}$ where $s$ is a non-negative real 
number. Summing over $n$ and using the {\em generating function} 
\eqa
g(u,s)={\exp}(su)J_0(s{\sqrt{1-u^2}})= \sum_{n=0}^{\infty}
{\frac{1}{n!}}P_n(u)s^n,
\ena
for Legendre polynomials [8], we get the continuous spectrum of solutions
\eqa
{\Phi}_s(u){\ }{\propto}{\ }{\exp}(-s) \sum_{n=0}^{\infty}{\frac{1}{n!}}
P_n(u)s^n={\exp}{\Big (}-2s{\frac{{\xi}^2}{{\Xi}_0^2}}{\Big )}
J_0{\Big (}2s{\frac{{\xi}}{{\Xi}_0}}
{\sqrt{1-{\frac{{\xi}^2}{{\Xi}_0^2}}}}{\Big )}.
\ena
Setting $k{\equiv}{\frac{2s}{{\Xi}_0}}$ and then taking the limit 
${\Xi}{\rightarrow}{\infty}$ in equation (21), we get back the Newtonian case
(with $G^{\rm S}{\approx}G_{\rm N}$, where $G_{\rm N}$ is Newton's constant
measured at epoch $t_0$).

To find the mode surface densities ${\bar {\Sigma}}_n(u)$ and  
${\bar {\Sigma}}_n^{\rm i}(u)$ corresponding to the specific mode solutions 
$-C_nP_n(u)$ and $-C_n^{\rm i}Q_n(u)$, respectively, we lay a Gauss surface 
around the disk and use Gauss' theorem across it. This procedure is exactly 
similar to the Newtonian case treated in [7]. Assuming a weak field 
(${\bar B}_n{\equiv}1+{\frac{2}{c^2}}{\Phi}_n{\approx}1$) we find
\eqa
{\bar {\Sigma}}_n(u)={\frac{c^2C_n}{2{\pi}G^{\rm S}{\Xi}_0}}(2n+1)P_n(u), 
\qquad {\bar {\Sigma}}_n^{\rm i}(u)=
{\frac{c^2C_n^{\rm i}}{2{\pi}G^{\rm S}{\Xi}_0}}(2n+1)Q_n(u).
\ena
Due to the fact that the set of Legendre polynomials $P_n(u)$, $u{\in}(-1,1)$,
is complete and orthogonal [9], it is possible to expand any real surface 
density ${\bar {\Sigma}}(u)$ in terms of the mode surface densities
${\bar {\Sigma}}_n(u)$. Setting $C_n{\equiv}CS_n$ (where $C$ is some nonzero
constant) and summing over $n$, we find
\eqa
{\bar {\Sigma}}(u)= \sum_{n=0}^{\infty}{\bar {\Sigma}}_n(u)=
{\frac{c^2C}{{\pi}G^{\rm S}{\Xi}_0}} \sum_{n=0}^{\infty}
{\frac{1}{2}}(2n+1)S_nP_n(u).
\ena
But this means that ${\bar {\Sigma}}(u)$ is expressed as a {\em Legendre
Fourier series} [9], and that its inverse $S_n$ is the {\em finite Legendre
transform} [9] of ${\frac{{\pi}G^{\rm S}{\Xi}_0}{c^2C}}{\bar {\Sigma}}(u)$, i.e.,
\eqa
S_n={\frac{{\pi}G^{\rm S}{\Xi}_0}{c^2C}} \int_{-1}^1
{\bar {\Sigma}}(u')P_n(u')du',
\ena
so that
\eqa
{\bar {\Sigma}}(u)={\frac{1}{2}} \sum_{n=0}^{\infty}(2n+1)P_n(u)
 \int_{-1}^1{\bar {\Sigma}}(u')P_n(u')du'.
\ena
The real solution ${\Phi}(u)$ corresponding to the given surface density
${\bar {\Sigma}}(u)$ is then obtained by summing up the mode solutions 
$-C_nP_n(u)$, i.e.
\eqa
{\Phi}(u)=-c^2C \sum_{n=0}^{\infty} S_nP_n(u)=-{\pi}G^{\rm S}{\Xi}_0 
\sum_{n=0}^{\infty}P_n(u) \int_{-1}^1{\bar {\Sigma}}(u')P_n(u')du'.
\ena
The corresponding series solution ${\bar B}_{\rm {real}}(u,z)$ is then found by 
combining equations (15), (19), (25) and (26). The result is
\eqa
{\bar B}_{\rm {real}}(u,z)&=&1+{\frac{2}{c^2}}{\Phi}(u)
+{\frac{4{\pi}G^{\rm S}}{c^2}}|z|{\bar {\Sigma}}(u) \nonumber \\
&&-{\frac{4{\pi}G^{\rm S}}{c^2{\Xi}_0}}z^2\sum_{n=0}^{\infty}n(n+1)P_n(u)
\int_{-1}^1{\bar {\Sigma}}(u')P_n(u')du'+{\cdots}.
\ena
The circular speed ${\bar w}_{\rm {real}}$ due to the solution (26) as 
function of radius of the disk can now be found from the usual non-relativistic 
formula (this is justified for a weak field ${\bar B}_{\rm {real}}{\approx}1$, 
see [4]). Expressed as a function of $u$ we find
\eqa
{\bar w}^2_{\rm {real}}(u)&=&2{\pi}G^{\rm S}{\Xi}_0(1-u) \sum_{n=0}^{\infty}
{\frac{{\partial}P_n(u)}{{\partial}u}} \int_{-1}^1{\bar {\Sigma}}(u')P_n(u')du'
\nonumber \\
{\!}{\!}{\!}{\!}{\!}{\!}{\!}
&=&{\frac{2{\pi}G^{\rm S}{\Xi}_0}{1+u}} \sum_{n=0}^{\infty}(n+1){\Big (}uP_n(u)
-P_{n+1}(u){\Big )} \int_{-1}^1{\bar {\Sigma}}(u')P_n(u')du'.
\ena
It cannot be expected that rotation curves found from equation (28) should 
deviate significantly from their Newtonian counterparts (they don't). Thus it
would seem that the need for galactic DM is the same as for the standard model.
However, here we have not taken into account possible solutions constructed 
from the specific mode solutions $-C_n^{\rm i}Q_n(u)$. As we shall see later, if 
we do this, new possibilities of getting rid of galactic DM open up. But 
first, in the next section, we need to consider restrictions coming from 
boundary conditions. 
\subsection{Boundary conditions}
So far we have implicitly assumed that there are no particular preferences 
regarding the form of ${\bar {\Sigma}}(u)$ as long as it is physically 
reasonable and the resulting ${\Phi}(u)$ is small everywhere. But is this true? 
To answer that question, we first notice from equation (26) that the real 
potential at the centre of the disk is given by
\eqa
{\Phi}(1)=-{\pi}G^{\rm S}{\Xi}_0 \sum_{n=0}^{\infty} \int_{-1}^1
{\bar {\Sigma}}(u')P_n(u')du'=-{\frac{{\pi}G^{\rm S}{\Xi}_0}{{\sqrt{2}}}}
 \int_{-1}^1{\frac{{\bar {\Sigma}}(u')du'}{{\sqrt{1-u'}}}},
\ena
where the last expression follows from the generating function [8]
\eqa
\sum_{m=0}^{\infty}P_m(u)s^m={\frac{1}{\sqrt{1-2us+s^2}}}, \qquad -1<s<1,
\ena
for the borderline case $s=1$. Moreover, equation (29) may be written in the
form
\eqa
{\Phi}(1)=-c^2{\frac{{\bar {\Sigma}}_{\rm +}}{{\bar {\Sigma}}_*}}, \qquad
{\bar {\Sigma}}_{\rm +}{\equiv}{\frac{1}{2{\sqrt 2}}}
\int_{-1}^1{\frac{{\bar {\Sigma}}(u')du'}{{\sqrt{1-u'}}}}, \qquad
{\bar {\Sigma}}_*{\equiv}{\frac{c^2}{2{\pi}G^{\rm S}{\Xi}_0}},
\ena
where ${\bar {\Sigma}}_{\rm +}$ is a weighted average, and where the constant 
${\bar {\Sigma}}_*$ sets a specific surface density scale depending on the 
finite size of space. Besides, a second weighted average surface density
${\bar {\Sigma}}_{\rm -}$, related to the total (active) mass $M_{t_0}$ of the 
disk, can be defined by
\eqa
{\bar {\Sigma}}_{\rm -}{\equiv}{\frac{1}{2{\sqrt 2}}}
\int_{-1}^1{\frac{{\bar {\Sigma}}(u')du'}{{\sqrt{1+u'}}}}=
{\frac{M_{t_0}}{2{\pi}{\Xi}_0^2}}, \qquad \Rightarrow \qquad
{\Phi}(-1)=-c^2{\frac{{\bar {\Sigma}}_{\rm -}}{{\bar {\Sigma}}_*}}
=-{\frac{G^{\rm S}M_{t_0}}{{\Xi}_0}},
\ena
where ${\Phi}(-1)$ is found from an expression similar to equation (29) by
using equation (30) for the borderline case $s=-1$.
We notice that ${\bar {\Sigma}}_*$ is a purely geometric quantity, whereas 
${\bar {\Sigma}}_{\rm +}$ and ${\bar {\Sigma}}_{\rm -}$ depend on the real surface
density profile. 

Next we notice that ${\sqrt {{\mid}{\Phi}(-1){\mid}}}$ represents a specific 
(non-vanishing) velocity scale.
This indicates the possibility of defining some quantity
${\frac{\pi}{\sqrt 2}}[{\bar {\sigma}}(-1)-{\bar {\sigma}}(1)]
{\sim}{\bar {\Sigma}}_*$ with the property that it relates ${\Phi}(1)$ to 
${\sqrt {{\mid}{\Phi}(-1){\mid}}}$ via a definition similar to equation (31).
By combining equations (31) and (32) such a relationship may readily be found.
That is, we may define
\eqa
{\Phi}(1)=-c{\sqrt {{\mid}{\Phi}(-1){\mid}{\frac{{\bar {\Sigma}}_*}
{{\bar {\Sigma}}_{\rm -}}}}}{\frac{{\bar {\Sigma}}_{\rm +}}{{\bar {\Sigma}}_*}}
{\equiv}-c{\sqrt {{\mid}{\Phi}(-1){\mid}}}{\frac{\pi}{\sqrt 2}}
{\frac{[{\bar {\sigma}}(-1)-{\bar {\sigma}}(1)]}{{\bar {\Sigma}}_*}}.
\ena
Note that equation (33) involves the geometric quantity 
${\bar {\Sigma}}_*$, but since ${\Phi}(1)={\frac{{\bar {\Sigma}}_{\rm +}}
{{\bar {\Sigma}}_{\rm -}}}{\Phi}(-1)$, the analogous relationship between 
${\Phi}(1)$ and ${\Phi}(-1)$ does not. Also note that the factor 
${\pi}/{\sqrt{2}}$ is included into the definition (33) since 
${\bar {\sigma}}(-1)-{\bar {\sigma}}(1)$ should be more similar to a mode 
surface density (see equation (22)) rather than to a weighted average like 
${\bar {\Sigma}}_+$.

A definition similar to equation (33) may be made for the contribution 
${\Phi}_{{\geq}u}(1)$ to ${\Phi}(1)$ from the part of the disk interior to some 
arbitrary coordinate $u$. The purpose of such a definition is to construct a 
new ``associated'' surface density ${\bar {\sigma}}(u)$. That is, we may define
\eqa
{\Phi}_{{\geq}u}(1){\equiv}-c{\sqrt {{\mid}{\Phi}(-1){\mid}}}
{\frac{\pi}{\sqrt 2}}
{\frac{[{\bar {\sigma}}(u)-{\bar {\sigma}}(1)]}{{\bar {\Sigma}}_*}}, \qquad
{\Phi}_{{\geq}u}(1){\equiv}-{\frac{{\pi}G^{\rm S}{\Xi}_0}{{\sqrt{2}}}}
 \int_{u}^1{\frac{{\bar {\Sigma}}(u')du'}{{\sqrt{1-u'}}}},
\ena
or equivalently (where the constant ${\bar {\sigma}}(1)$ must be determined 
separately, see below)
\eqa
{\bar {\sigma}}(u){\equiv}{\bar {\sigma}}(1)+{\frac{1}{2{\pi}}}
{\sqrt {\frac{{\bar {\Sigma}}_{*}}{{\bar {\Sigma}}_{\rm -}}}}
\int_{u}^1{\frac{{\bar {\Sigma}}(u')du'}{{\sqrt{1-u'}}}}, \qquad
{\bar {\sigma}}(-1){\equiv}{\bar {\sigma}}(1)+{\frac{\sqrt 2}{{\pi}}}
{\sqrt {\frac{{\bar {\Sigma}}_{*}}{{\bar {\Sigma}}_{\rm -}}}}
{\bar {\Sigma}}_{\rm +}.
\ena 
We see from the definition (35) that ${\bar {\sigma}}(u)$ is increasing 
from the centre of the disk and outwards. Thus, ${\bar {\sigma}}(u)$ could be 
interpreted as some kind of ``inverted'' surface density. This means that
${\bar {\sigma}}(u)$ should not be considered as an independent, gravitating 
source. Rather, ${\bar {\sigma}}(u)$ should give some restrictions on the 
possible forms of ${\bar {\Sigma}}(u)$. Such restrictions can be found
by requiring that ${\bar {\sigma}}(u)$ should be linearly related to 
${\bar {\Sigma}}(u)$, so that the corresponding potential can be written as a 
linear combination of ${\Phi}(u)$ and some constant potential. Then
${\bar {\sigma}}(u)$ is not independent. A ``preferred'' form of 
${\bar {\sigma}}(u)$ can thus be found by requiring that 
${\bar {\sigma}}(u)-{\bar {\sigma}}(1)+{\lambda}{\bar {\Sigma}}(u)=
{\bar {\sigma}}(-1)$, where ${\lambda}$ is a constant, or equivalently
\eqa
{\bar {\sigma}}(u)={\bar {\sigma}}(-1)-{\frac{{\bar {\sigma}}(-1)}
{{\bar {\Sigma}}(1)}}{\Big [}{\bar {\Sigma}}(u)-{\bar {\Sigma}}(-1){\Big ]},
\quad \Rightarrow \quad
{\bar {\sigma}}(1)={\bar {\Sigma}}(-1){\frac{{\bar {\sigma}}(-1)}
{{\bar {\Sigma}}(1)}}.
\ena
(We see that in this case, since ${\bar {\Sigma}}(-1)$ should be negligible,
${\bar {\sigma}}(1)$ should be also.) Equation (36) is then an integral 
equation determining the most basic form of the real surface density 
${\bar {\Sigma}}(u)$ for an isolated disk. To find exactly what this form is, 
it is convenient to turn equation (36) into a first order separable 
differential equation by taking the derivative w.r.t. $u$ at both sides of it. 
Solving this equation is straightforward, and the result is an exponential 
disk, i.e.,
\eqa
{\bar {\Sigma}}(u)={\bar {\Sigma}}(1){\exp}{\Big [}
-{\Big (}{\frac{{\bar {\Sigma}}(1)}{{\bar {\Sigma}}_+}}-
{\frac{{\bar {\Sigma}}(-1)}{{\bar {\Sigma}}_+}}{\Big )}{\sqrt{(1-u)/2}}{\Big ]}
{\equiv}{\bar {\Sigma}}(1){\exp}{\Big [}
-{\frac{{\Xi}_0}{{\xi}_{\rm d}}}{\sqrt{(1-u)/2}}{\Big ]},
\ena
where ${\xi}_{\rm d}{\equiv}{\Xi}_0{\bar {\Sigma}}_+/
[{\bar {\Sigma}}(1)-{\bar {\Sigma}}(-1)]$ is the disk length (at epoch $t_0$).
This result answers the question we posed at the beginning of this section;
the simple requirement that ${\bar {\Sigma}}(u)$ and ${\bar {\sigma}}(u)$ 
should be linearly related implies that there is a particular preference 
regarding the form of ${\bar {\Sigma}}(u)$. That is, it would seem that the 
exponential disk should represent a preferred surface density profile among all
the possibilities that might exist. This is confirmed observationally, since an 
exponential surface density is the hallmark density profile of the outer 
regions of spiral galaxies. We will return to the exponential disk in 
section 4.
\subsection{The induced solution}
Contrary to the Legendre polynomials $P_n(u)$, the functions $Q_n(u)$ are not
polynomials, and nor do they constitute an orthogonal set for $u{\in}(-1,1)$.
Rather, the functions $Q_n(u)$ can be separated into two subsets depending on 
whether $n$ is even or odd. That is, each function with odd $n$ is orthogonal 
to every function with even $n$ and {\em vice versa}. On the other hand, 
functions within each subset are linearly dependent. This can be easily seen 
from the formulae [8]
\eqa
\int_{-1}^1Q_n(u)Q_n(u)du&=&{\frac{{\frac{{\pi}^2}{6}}+
2\sum_{k=1}^n{\frac{1}{k^2}}}{2n+1}}, \nonumber \\
\int_{-1}^1Q_n(u)Q_m(u)du&=&{\frac{1+(-1)^{n-m}}{(m-n)(m+n+1)}}{\Big (}
\sum_{k=1}^n-\sum_{k=1}^m{\Big )}{\frac{1}{k}}, \qquad n{\neq}m.
\ena
The problem now is to construct a solution ${\Phi}^{\rm i}(u)$ from the mode 
solutions $-C^{\rm i}_nQ_n(u)$ such that ${\Phi}(u)$ and ${\Phi}^{\rm i}(u)$ are 
linearly independent, i.e., we require that $\int_{-1}^1{\Phi}^{\rm i}(u){\Phi}
(u)du=0$. However, since there are only two sets of linearly independent mode 
solutions, it must be possible to find many such solutions by summing over 
different numbers of mode solutions (in general, at least two mode solutions 
must be included, one from each linearly independent set). This means that, 
merely requiring linear independence is not sufficient to arrive at a unique 
solution ${\Phi}^{\rm i}(u)$. However, a unique linearly independent solution 
${\Phi}^{\rm i}(u)$ can indeed be found by summing over {\em all} the mode 
solutions. We will call this solution {\em the induced solution}, since it is 
found indirectly by summing up all the mode solutions $C^{\rm i}_nQ_n(u)$ such 
that every term in the mode sum has a linearly independent counterpart 
$C_nP_n(u)$ from equation (26). Moreover, the requirement of linear 
independence is not trivial since it forces the constants $C^{\rm i}_n$ to be 
dependent on $S_n$ and thus the real surface density. The corresponding surface
density ${\bar{\Sigma}}^{\rm i}(u)$ will be called {\em the induced density}. 
The induced density is not real, but could still have physical consequences 
indirectly.

The induced solution ${\Phi}^{\rm i}(u)$ obtained by summing over all the mode 
solutions can be found by assuming that the coefficients $C^{\rm i}_n$ can be 
written in the form $C^{\rm i}_n=C^{\rm i}S_n$, where $C^{\rm i}$ is a normalisation 
constant. Then, using the formula [8]
\eqa
\int_{-1}^1P_n(u)Q_m(u)du={\frac{1-(-1)^{n+m}}{(n-m)(m+n+1)}}, \qquad 
n{\neq}m,
\ena
it is easy to see that
\eqa
\int_{-1}^1{\Phi}^{\rm i}(u){\Phi}(u)du{\ }{\propto}{\ }\sum_{n=0}^{\infty}
\sum_{\scriptstyle{m=0}\atop\scriptstyle{m{\neq}n}}^{\infty}
S_nS_m{\frac{1-(-1)^{n+m}}{(n-m)(m+n+1)}}=0,
\ena
from the obvious antisymmetry obtained by permuting the summation indices. To 
completely specify the solution ${\Phi}^{\rm i}(u)$, it remains to specify 
the normalisation constant $C^{\rm i}$. But the only natural choice is really to 
set $C^{\rm i}=C$. We thus have that
\eqa
C^{\rm i}_n=C_n=CS_n={\frac{{\pi}G^{\rm S}{\Xi}_0}{c^2}}
{\int_{-1}^1{\bar {\Sigma}}(u')P_n(u')du'}. 
\ena
The solution ${\Phi}^{\rm i}(u)$ is now completely specified, and we have that
\eqa
{\Phi}^{\rm i}(u)=-c^2C \sum_{n=0}^{\infty}S_nQ_n(u)=-{\pi}G^{\rm S}{\Xi}_0
 \sum_{n=0}^{\infty} \int_{-1}^1{\bar {\Sigma}}(u')P_n(u')du'Q_n(u).
\ena
Equation (42) may be written in a more convenient form by using the identity 
[8] (the integral being defined by its Cauchy principal value)
\eqa
Q_n(u)={\frac{1}{2}}\int_{-1}^1{\frac{P_n(s)ds}{u-s}},
\ena
so that by using equation (26) we find that (again using Cauchy principal 
values)
\eqa
{\Phi}^{\rm i}(u)={\frac{1}{2}} \int_{-1}^1{\frac{{\Phi}(s)ds}{u-s}}.
\ena
From equations (22), (25), (41) and (43) we can also find the induced density
${\bar {\Sigma}}^{\rm i}(u)$ corresponding to the induced solution 
${\Phi}^{\rm i}(u)$, i.e.,
\eqa
{\bar {\Sigma}}^{\rm i}(u)=
\sum_{n=0}^{\infty}{\frac{c^2C_n}{2{\pi}G^{\rm S}{\Xi}_0}}(2n+1)Q_n(u)=
{\frac{1}{2}} \int_{-1}^1{\frac{{\bar {\Sigma}}(s)ds}{u-s}}.
\ena
However, there is a fundamental problem with the induced quantities. That is,
both ${\Phi}^{\rm i}(u)$ and ${\bar{\Sigma}}^{\rm i}(u)$ will in general contain 
generic logarithmic divergences (typically located at $u={\pm}1$) inherited 
from the mode solutions $Q_n(u)$. This means that ${\Phi}^{\rm i}(u)$ and 
${\bar{\Sigma}}^{\rm i}(u)$ cannot be used directly as physical quantities. But 
${\bar{\Sigma}}^{\rm i}(u)$ may be used {\em indirectly}, since it is possible 
to construct a new, geometric, singularity-free surface density from it. We 
will show how to do this in the next section.
\subsection{The associated induced potential}
Although ${\bar{\Sigma}}^{\rm i}(u)$ cannot be used directly, it still may have
physical significance. The reason for this is that it is possible to construct
a new surface density ${\bar{\sigma}}^{\rm i}(u)$ from ${\bar{\Sigma}}^{\rm i}(u)$ 
using the results found in section 3.2. There, the definition (35) of 
${\bar{\sigma}}(u)$ was motivated from the possibility of restricting possible
forms of ${\bar{\Sigma}}(u)$ due to boundary conditions. However, since 
${\bar{\sigma}}(u)$ is by construction unphysical, there never was any reason 
to interpret it as an independent gravitating source. On the other hand, there 
is always the possibility that a definition similar to (35), but with
${\bar{\Sigma}}^{\rm i}(u')$ substituted for ${\bar{\Sigma}}(u')$, may have the
properties that makes it possible to use it as a gravitating, geometric source. 
That is, we may construct the so-called
{\em associated induced surface density} ${\bar{\sigma}}^{\rm i}(u)$, defined by
\eqa
{\bar{\sigma}}^{\rm i}(u){\equiv}{\bar{\sigma}}^{\rm i}(1)+{\frac{1}{2{\pi}}}
{\sqrt {\frac{{\bar {\Sigma}}_{*}}{{\bar {\Sigma}}_{\rm -}}}}
\int_{u}^1{\frac{{\bar {\Sigma}}^{\rm i}(u')du'}{{\sqrt{1-u'}}}}, \qquad
{\bar{\sigma}}^{\rm i}(1){\approx}0,
\qquad
{\frac{{\bar {\Sigma}}_{*}}{{\bar {\Sigma}}_{\rm -}}}=
{\frac{c^2{\Xi}_0}{M_{t_0}G^{\rm S}}}.
\ena
Note that, by integrating over ${\bar{\Sigma}}^{\rm i}(u')$, its generic 
logarithmic divergences disappear, so that ${\bar{\sigma}}^{\rm i}(u)$ does not 
contain such singularities. Moreover, since it is possible that 
${\bar{\Sigma}}^{\rm i}(u)$ may change sign somewhere in the interval 
$u{\in}(-1,1)$, ${\bar{\sigma}}^{\rm i}(u)$ may take a form similar to some 
physical surface density profile. Finally, unlike ${\bar{\sigma}}(u)$, 
${\bar{\sigma}}^{\rm i}(u)$ can certainly not be algebraically related to 
${\bar{\Sigma}}(u)$. Thus ${\bar{\sigma}}^{\rm i}(u)$ may be considered as a 
physical, independent gravitating {\em geometric} quantity, but not as a 
material density.

We may now use ${\bar{\sigma}}^{\rm i}(u)$ as a source in equation (26) to get a 
new potential, i.e., the so-called {\em associated induced potential} 
${\Psi}(u)$ defined by
\eqa
{\Psi}(u)=-{\pi}G^{\rm S}{\Xi}_0 \sum_{n=0}^{\infty}
P_n(u) \int_{-1}^1{\bar {\sigma}}^{\rm i}(u')P_n(u')du'.
\ena
Since ${\bar{\sigma}}^{\rm i}(u)$ does not contain generic divergences, we see 
from equation (47) that ${\Psi}(u)$ should be non-singular everywhere, so it 
may be accepted as a physical quantity. Thus to any real surface density 
profile ${\bar{\Sigma}}(u)$ and its corresponding real potential ${\Phi}(u)$, 
there is always associated a surface density ${\bar{\sigma}}^{\rm i}(u)$ 
(playing the role of galactic ``dark matter''), and its corresponding potential 
${\Psi}(u)$. These quantities should always be considered together when making 
predictions. 

The series solution ${\bar B}(u,z)$ containing both real and associated induced
contributions is then given by
\eqa
{\bar B}(u,z)&=&1+{\frac{2}{c^2}}{\Big (}{\Phi}(u)+{\Psi}(u){\Big )}
+{\frac{4{\pi}G^{\rm S}}{c^2}}|z|{\Big (}{\bar {\Sigma}}(u)+
{\bar {\sigma}}^{\rm i}(u){\Big )} \nonumber \\
&-&{\frac{4{\pi}G^{\rm S}}{c^2{\Xi}_0}}z^2\sum_{n=0}^{\infty}n(n+1)P_n(u)
 \int_{-1}^1{\Big (}{\bar {\Sigma}}(u')+{\bar{\sigma}}^{\rm i}(u'){\Big )}
P_n(u')du'+{\cdots}.
\ena
Note that the second order term in equation (48) is divergent for $u=1$, 
$z{\neq}0$. See the next section for more comments. Moreover, the 
(non-relativistic) circular speed ${\bar w}_{\rm circ}$ calculated from equation 
(48) is due to both real matter and associated induced ``phantom'' matter, and 
similarly to equation (28) we find that
\eqa
{\bar w}^2_{\rm {circ}}(u)
={\frac{2{\pi}G^{\rm S}{\Xi}_0}{1+u}} \sum_{n=0}^{\infty}(n+1){\Big (}uP_n(u)-
P_{n+1}(u){\Big )} \int_{-1}^1{\Big (}{\bar {\Sigma}}(u')
+{\bar{\sigma}}^{\rm i}(u){\Big )}P_n(u')du'.
\ena
We see from equations (44) and (45) that the induced solutions 
and densities are obtained by integrating real solutions 
and densities over the whole disk; thus these quantities describe non-local, 
{\em collective} properties of the system. This is also true for
${\bar{\sigma}}^{\rm i}(u)$. That is why any extra gravitational 
acceleration obtained from ${\Psi}(u)$ should not be interpreted 
as a fundamental modification of the Newtonian force law such as in MOND - 
rather the extra acceleration should be seen as an emergent property of the 
whole system.
\subsection{An important transformation}
Given as input the real surface density ${\bar {\Sigma}}(u)$, one should now in
principle be able to calculate the real potential ${\Phi}(u)$, the real series 
solution ${\bar B}_{\rm real}(u,z)$, the corresponding induced quantities 
${\bar {\Sigma}}^{\rm i}(u)$ and ${\Phi}^{\rm i}(u)$, the associated induced
quantities ${\bar{\sigma}}^{\rm i}(u)$ and ${\Psi}(u)$, the total series 
solution ${\bar B}(u,z)$, and finally the rotation curve from equation (49). 
However, as seen from equations (26) and (47), the expressions for ${\Phi}(u)$ 
and ${\Psi}(u)$ contain an infinite sum over (a product of) Legendre 
polynomials. Since this sum will in general converge slowly, its presence makes
numerical calculations quite awkward. Fortunately, it is possible to rewrite 
this infinite sum in terms of an elliptic integral, making numerical 
calculations much easier. The key to this important transformation is using 
the generating function [10] (${\mid}s{\mid}<1$)
\eqa
\sum_{m=0}^{\infty} P_m(u)P_m(u')s^m={\frac{1}{\pi}}
\int_{0}^{\pi}{\frac{d{\omega}}{{\sqrt{1-2s{\Big (}uu'+
{\sqrt{(1-u'^2)(1-u^2)}}{\cos}{\omega}{\Big )}+s^2}}}},
\ena
for the borderline case $s=1$. (Note that, for the special case $u'=1$, we get
back equation (30).)

We now use equation (50) to rewrite equations (26) and (47). Interchanging the 
sum and the integral in these equations, adding them and then using equation 
(50), we get
\eqa
{\Phi}(u)+{\Psi}(u)&=&-{\frac{G^{\rm S}{\Xi}_0}{\sqrt{2}}}  \int_{0}^{\pi}
 \int_{-1}^1{\frac{{\Big [}{\bar {\Sigma}}(u')+{\bar {\sigma}}^{\rm i}(u')
{\Big ]}du'd{\omega}}{{\sqrt{1-uu'-{\sqrt{(1-u'^2)(1-u^2)}}{\cos}{\omega}}}}} 
\nonumber \\
&=&-{\sqrt{2}}G^{\rm S}{\Xi}_0 \int_{-1}^1{\frac{{\Big [}{\bar {\Sigma}}(u')+
{\bar {\sigma}}^{\rm i}(u'){\Big ]}K{\Big (}
{\sqrt{{\frac{2{\sqrt{(1-u^2)(1-u'^2)}}}
{1-uu'+ {\sqrt{(1-u^2)(1-u'^2)}}}}}}{\Big )}du'}
{{\sqrt{1-uu'+{\sqrt{(1-u'^2)(1-u^2)}}}}}} \nonumber
\ena
\vspace*{-5mm}
\eqa
=-2{\sqrt{2}}G^{\rm S}{\Xi}_0 \int_{-1}^1{\frac{{\Big [}{\bar {\Sigma}}(u')+
{\bar {\sigma}}^{\rm i}(u'){\Big ]}K{\Big (}{\frac{{\sqrt{(1-u^2)(1-u'^2)}}}
{1-uu'+ {\mid}u-u'{\mid}}}{\Big )}du'}
{{\sqrt{1-uu'+{\sqrt{(1-u'^2)(1-u^2)}}}}+
{\sqrt{1-uu'-{\sqrt{(1-u'^2)(1-u^2)}}}}}},
\ena
where $K(k){\equiv} \int_0^{{\pi}/2}d{\theta}/{\sqrt{1-k^2{\sin}^2{\theta}}}$ is 
the complete elliptic integral of the first kind [8]. The last form (see [10]) 
of equation (51) is a little better to use for numerical purposes. 

Similarly, it may also be tempting to interchange the infinite sum and the 
integral present in the second-order term of equation (48), and then transform 
the infinite sum into an integral. However, as we shall see, this procedure 
does not work for higher-order terms, since all will be divergent. To 
illustrate this, a somewhat lengthy calculation using equation (50) yields
\eqa
\sum_{m=0}^{\infty}m(m+1)P_m(u)P_m(u')= \lim_{s{\rightarrow}1}
{\Big [}{\frac{\partial}{{\partial}s}}s^2{\frac{\partial}{{\partial}s}}
\sum_{m=0}^{\infty} P_m(u)P_m(u')s^m{\Big ]} \nonumber \\
={\frac{E(k)-(1-uu'-{\sqrt{(1-u^2)(1-u'^2)}})K(k)}
{2{\sqrt{2}}{\pi}{\mid}u-u'{\mid}{\sqrt{1-uu'-{\sqrt{(1-u'^2)(1-u^2)}}}}}},
\quad k^2{\equiv}{\frac{2{\sqrt{(1-u^2)(1-u'^2)}}}
{1-uu'+ {\sqrt{(1-u^2)(1-u'^2)}}}},
\ena
where $E(k){\equiv} \int_0^{{\pi}/2}{\sqrt{1-k^2{\sin}^2{\theta}}}d{\theta}$ is
the complete elliptic integral of the second kind [8]. To get an expression for
${\bar B}(u,z)$ (for small ${\mid}z{\mid}$) more suitable for numerical 
calculations, one may now insert equation (51) into equation (48). However, if 
one tries to insert equation (52) into the quadratic term, interchanging the 
infinite sum and the integral, it is straightforward to see that this term will
diverge. This behaviour is quite similar to the Newtonian case (where the 
potential takes the form of a double improper integral [7]) if one tries to 
expand the Newtonian potential in powers of ${\mid}z{\mid}$; interchanging the 
two improper integrals implies that all terms with power ${\geq}2$ will 
diverge, yet if all the terms are included, the resulting exact result is 
finite. But only including the linear term in the series expansion approximates
the exact result well for small enough ${\mid}z{\mid}$. One expects that this 
is valid for equation (48) also, so that skipping terms with power ${\geq 2}$ 
is justified for small ${\mid}z{\mid}$. Note that in equation (48)
(also just as for its Newtonian counterpart), all terms of order ${\geq}2$ will 
diverge on the $z$-axis for $z{\neq}0$.

Of course, if ${\mid}z{\mid}$ is not small enough, the second order term (and 
possibly higher-order terms) must be calculated in some admissible way and 
included. That is, it would be preferable to find some other way of rewriting 
the series expansion in ${\mid}z{\mid}$ (i.e., without using equation (52)), 
such that all terms can be expressed in a form suitable for numerical 
calculations. Fortunately, such a form for the second order term can be found 
from equations (48) and (13) (with ${\bar V}=0$). From these equations we find 
that (for small ${\mid}z{\mid}$)
\eqa
{\bar B},_{zz}&=&-{\frac{2}{c^2}}{\Big [}
(1-{\frac{{\xi}^2}{{\Xi}_0^2}})({\Phi},_{{\xi}{\xi}}+{\Psi},_{{\xi}{\xi}})+
{\frac{1}{\xi}}(1-{\frac{3{\xi}^2}{{\Xi}_0^2}})({\Phi},_{\xi}+{\Psi},_{\xi})
{\Big ]}+O({\mid}z{\mid}) \nonumber \\
&=&-{\frac{8}{c^2{\Xi}_0^2}}{\Big [}(1+u)({\Phi},_{uu}+{\Psi},_{uu})
-2u({\Phi},_u+{\Psi},_u){\Big ]}+O({\mid}z{\mid}),
\ena
so that the first few terms of the rewritten series expansion read
\eqa
{\bar B}(u,z)&=&1+{\frac{2}{c^2}}{\Big (}{\Phi}(u)+{\Psi}(u){\Big )}
+{\frac{4{\pi}G^{\rm S}}{c^2}}|z|{\Big (}{\bar {\Sigma}}(u)+
{\bar {\sigma}}^{\rm i}(u){\Big )} \nonumber \\
&&-{\frac{4z^2}{c^2{\Xi}_0^2}}{\Big [}(1+u)({\Phi},_{uu}+{\Psi},_{uu})
-2u({\Phi},_u+{\Psi},_u){\Big ]}+O({\mid}z^3{\mid}).
\ena
Note that the second order and higher-order terms are still expected to 
diverge in the limit $u{\rightarrow}1$, $z{\neq}0$. This means that, to
do calculations to a given accuracy at some fixed $z{\neq}0$, one needs
to include ever more higher-order terms into the series expansion when moving
towards the $z$-axis. 

Finally, interchanging the sum and the integral in equation (49), yields, after
some tedious calculations, that
\eqa
{\bar w}^2_{\rm circ}(u)=
{\sqrt{2}}G^{\rm S}{\Xi}_0{\Bigg \{}{\frac{u}{1+u}} \int_{-1}^1
{\frac{{\Big [}{\bar {\Sigma}}(u')+{\bar {\sigma}}^{\rm i}(u'){\Big ]}
K{\Big (}{\sqrt{{\frac{2{\sqrt{(1-u^2)(1-u'^2)}}}
{1-uu'+ {\sqrt{(1-u^2)(1-u'^2)}}}}}}{\Big )}du'}
{{\sqrt{1-uu'+{\sqrt{(1-u'^2)(1-u^2)}}}}}} \nonumber \\
-{\frac{1}{1+u}} \int_{-1}^1{\frac{(u-u'){\Big [}{\bar {\Sigma}}(u')+
{\bar {\sigma}}^{\rm i}(u'){\Big ]}E{\Big (}
{\sqrt{{\frac{2{\sqrt{(1-u^2)(1-u'^2)}}}
{1-uu'+ {\sqrt{(1-u^2)(1-u'^2)}}}}}}{\Big )}du'}
{{\mid}u-u'{\mid}{\sqrt{1-uu'-{\sqrt{(1-u'^2)(1-u^2)}}}}}}{\Bigg \}}.
\ena
Inserting any given surface density ${\bar {\Sigma}}(u)$ and its associated 
induced surface density ${\bar {\sigma}}^{\rm i}(u)$ into equation (55) now 
yields the full rotation curve of the disk.
\section{The exponential disk}
\subsection{Approximate solutions}
For spiral galaxies, the observed general trend is that surface brightness (and
thus the luminosity due to stars in the disk) falls off exponentially from the 
centre and outwards. Therefore, to explain spiral galaxy rotational curves 
without dark matter, one is required to assume that surface density profiles of
stars are proportional to luminosity profiles. This means that, in the general,
but idealised case of a truncated disk of extension ${\xi}_0$, we must have 
that
\eqa
 {\bar {\Sigma}}({\xi})= \left\{ \begin{array}{ll}
{\bar {\Sigma}}_{\rm c}{\exp}(-{\xi}/{\xi}_{\rm d}) & {\xi}{\leq}{\xi}_0, \\ 
0 & {\xi}>{\xi}_0, \end{array} \right. \qquad
{\bar {\Sigma}}(u)= \left\{ \begin{array}{ll}
{\bar {\Sigma}}_{\rm c}{\exp}{\Big (}-{\frac{{\Xi}_0}{{\xi}_{\rm d}}}
{\sqrt{{\frac{1-u}{2}}}}{\Big )} & 1{\geq}u{\geq}u_0, \\
0 & u<u_0, \end{array} \right.
\ena
where ${\bar {\Sigma}}_{\rm c}$ is the central surface density and ${\xi}_{\rm d}$ 
is the disk length (both taken at epoch $t_0$). Note that, since 
${\xi}_{\rm d}{\ll}{\Xi}_0$, ${\bar {\Sigma}}({\xi})$ falls off so fast that we 
are justified in treating spiral galaxies as truncated disks for sufficiently 
large ${\xi}_0$, i.e., for ${\xi}_0{\gg}{\xi}_{\rm d}$. For computational 
purposes, serious errors are hardly made if we choose ${\xi}_0$ to lie 
somewhere in the interval 20 ${\xi}_{\rm d}{\leq}{\xi}_0{\ll}{\Xi}_0$. On the 
other hand, truncation inevitably yields truncation singularities. However,
their contributions to calculations are usually negligible. In what
follows, we will assume that the disk is not truncated, but the results are
not significantly affected if it is.

Now we may calculate the induced surface density 
${\bar {\Sigma}}^{\rm i}({\xi})$ associated with the exponential disk. To do 
that, we compute the integral
\eqa
\int_{-1}^1(u-s)^{-1}{\exp}{\Big (}-{\frac{{\Xi}_0}{{\xi}_{\rm d}}}
{\sqrt{{\frac{1-s}{2}}}}{\Big )}ds \nonumber \\
={\exp}{\Big (}{\frac{{\Xi}_0}{{\xi}_{\rm d}}}{\sqrt{{\frac{1-u}{2}}}}{\Big )}
{\Big {\{}}{\rm Ei}{\Big [}-{\frac{{\Xi}_0}{{\xi}_{\rm d}}}{\Big (}
1+{\sqrt{{\frac{1-u}{2}}}}{\Big )}{\Big ]}-
{\rm Ei}{\Big [}-{\frac{{\Xi}_0}{{\xi}_{\rm d}}}
{\sqrt{{\frac{1-u}{2}}}}{\Big ]}{\Big {\}}} \nonumber \\
+{\exp}{\Big (}-{\frac{{\Xi}_0}{{\xi}_{\rm d}}}{\sqrt{{\frac{1-u}{2}}}}
{\Big )}{\Big {\{}}{\rm Ei}{\Big [}-{\frac{{\Xi}_0}{{\xi}_{\rm d}}}{\Big (}
1-{\sqrt{{\frac{1-u}{2}}}}{\Big )}{\Big ]}-
{\rm Ei}{\Big [}{\frac{{\Xi}_0}{{\xi}_{\rm d}}}
{\sqrt{{\frac{1-u}{2}}}}{\Big ]}{\Big {\}}} \nonumber \\
{\approx}
-{\exp}{\Big (}{\frac{{\Xi}_0}{{\xi}_{\rm d}}}{\sqrt{{\frac{1-u}{2}}}}{\Big )}
{\rm Ei}{\Big [}-{\frac{{\Xi}_0}{{\xi}_{\rm d}}}
{\sqrt{{\frac{1-u}{2}}}}{\Big ]}
-{\exp}{\Big (}-{\frac{{\Xi}_0}{{\xi}_{\rm d}}}{\sqrt{{\frac{1-u}{2}}}}
{\Big )}{\rm Ei}{\Big [}{\frac{{\Xi}_0}{{\xi}_{\rm d}}}
{\sqrt{{\frac{1-u}{2}}}}{\Big ]},
\ena
where the approximation holds as long as $u$ is not very close to $-1$, and if
${\xi}_{\rm d}{\ll}{\Xi}_0$. Here ${\rm Ei}(x)$ is the exponential integral 
defined by [8]
\eqa
{\rm Ei}(x){\equiv}- \int_{-x}^{\infty}{\frac{{\exp}(-s)}{s}}ds=
\int_{-{\infty}}^{x}{\frac{{\exp}(s)}{s}}ds, \quad 
{\rm Ei}(-x){\equiv}- \int_{x}^{\infty}{\frac{{\exp}(-s)}{s}}ds, \quad x>0.
\ena
We then find from equation (45) that
\eqa
{\bar {\Sigma}}^{\rm i}({\xi}){\approx}-{\frac{{\bar {\Sigma}}_{\rm c}}{2}}
{\Big {\{}}{\exp}({\xi}/{\xi}_{\rm d}){\rm Ei}(-{\xi}/{\xi}_{\rm d})+
{\exp}(-{\xi}/{\xi}_{\rm d}){\rm Ei}({\xi}/{\xi}_{\rm d}){\Big {\}}},
\qquad 0<{\xi}<{\Xi}_0.
\ena
Note that for (moderately) large distances, unlike ${\bar {\Sigma}}({\xi})$, 
${\bar {\Sigma}}^{\rm i}({\xi})$ tends slowly towards zero (from below) with 
increasing ${\xi}$. Besides, note that ${\bar {\Sigma}}^{\rm i}({\xi})$ is 
positive for small ${\xi}$ and that it diverges logarithmically towards the 
origin. (As can be seen from equation (57), the exact expression for
${\bar {\Sigma}}^{\rm i}({\xi})$ also has a similar singularity when 
${\xi}{\rightarrow}{\Xi}_0$.)

The associated induced surface density ${\bar {\sigma}}^{\rm i}({\xi})$ is now 
straightforwardly found from equations (46) and (59).
Neglecting the very small quantity ${\bar {\sigma}}^{\rm i}(0)$, we find that 
(setting $M_{t_0}{\approx}2{\pi}{\bar {\Sigma}}_{\rm c}{\xi}_{\rm d}^2$,
i.e., neglecting corrections coming from the finite size of space)
\eqa
{\bar {\sigma}}^{\rm i}({\xi}){\approx}{\frac{c}{2{\sqrt 2}{\pi}^2
{\xi}_{\rm d}}}{\sqrt{{\frac{M_{t_0}}{G^{\rm S}{\Xi}_0}}}}
{\Big {\{}}{\exp}(-{\xi}/{\xi}_{\rm d}){\rm Ei}({\xi}/{\xi}_{\rm d})-
{\exp}({\xi}/{\xi}_{\rm d}){\rm Ei}(-{\xi}/{\xi}_{\rm d}){\Big {\}}},
\qquad 0{\leq}{\xi}{\leq}{\Xi}_0.
\ena
Note that ${\bar {\sigma}}^{\rm i}({\xi})$ (even using the exact expression for  
${\bar {\Sigma}}^{\rm i}({\xi})$ found from equation (57)) is non-singular and 
non-negative everywhere. Thus ${\bar {\sigma}}^{\rm i}({\xi})$ may formally play 
the role of DM.

Next we find an approximative expression (valid if ${\xi}{\ll}{\Xi}_0$)
for the real potential ${\Phi}({\xi})$ from equation (51) (omitting 
${\bar {\sigma}}^{\rm i}(u)$). Splitting the integral up into two improper 
parts, we find that
\eqa
{\Phi}(\xi){\approx}
-4G^{\rm S}{\bar {\Sigma}}_{\rm c}{\xi}_{\rm d}{\Big [} \int_{0}^{{\frac{{\xi}}
{{\xi}_{\rm d}}}}+ \int_{{\frac{{\xi}}{{\xi}_{\rm d}}}}
^{{\frac{{\Xi}_0}{{\xi}_{\rm d}}}}{\Big ]}
{\frac{x{\exp}(-x)K{\Big (}2{\frac{{\sqrt{x{\xi}/{\xi}_{\rm d}}}}
{x+{\xi}/{\xi}_{\rm d}}}{\Big )}dx}
{x+{\xi}/{\xi}_{\rm d}}}, \qquad {\xi}{\ll}{\Xi}_0.
\ena
Note that, as can be readily verified numerically, 
${\Phi}({\xi})$ as calculated from equation (61) is essentially identical to 
its Newtonian counterpart ${\Phi}_{\rm N}({\xi})$ (valid for an infinite
exponential disk), given by [7]
\eqa
{\Phi}_{\rm N}({\xi})=-{\pi}G_{\rm N}{\bar {\Sigma}}_{\rm c}{\xi}{\Big [}I_0
({\xi}/2{\xi}_{\rm d})K_1({\xi}/2{\xi}_{\rm d})-I_1({\xi}/2{\xi}_{\rm d})
K_0({\xi}/2{\xi}_{\rm d}){\Big ]},
\ena
where $I_{\nu}(x)$ and $K_{\nu}(x)$ are modified Bessel functions [8]. Similarly,
the corresponding circular speed ${\bar w}_{\rm {real}}$ can be found 
approximately from equation (55) (omitting ${\bar {\sigma}}^{\rm i}(u)$), and 
this is also numerically very close to its Newtonian counterpart. 
That is, again splitting up integrals into improper parts, we find that 
for ${\xi}{\ll}{\Xi}_0$,
\eqa
{\bar w}^2_{\rm real}({\xi}){\approx}
2G^{\rm S}{\bar {\Sigma}}_{\rm c}{\xi}_{\rm d}{\Big [} \int_{0}^{{\frac{{\xi}}
{{\xi}_{\rm d}}}}+ \int_{{\frac{{\xi}}{{\xi}_{\rm d}}}}
^{{\frac{{\Xi}_0}{{\xi}_{\rm d}}}}{\Big ]}x{\exp}(-x){\Big [}
{\frac{K{\Big (}2{\frac{{\sqrt{x{\xi}/{\xi}_{\rm d}}}}
{x+{\xi}/{\xi}_{\rm d}}}{\Big )}}{x+{\xi}/{\xi}_{\rm d}}}-
{\frac{E{\Big (}2{\frac{{\sqrt{x{\xi}/{\xi}_{\rm d}}}}
{x+{\xi}/{\xi}_{\rm d}}}{\Big )}}
{x-{\xi}/{\xi}_{\rm d}}}{\Big ]}dx.
\ena
Note that the last term in equation (63) diverges for each integral, but that 
added together, the sum converges (taking its Cauchy principal value). 

An approximate expression for the associated induced potential ${\Psi}({\xi})$ 
is given by
\eqa
{\Psi}(\xi){\approx}-{\frac{{\sqrt 2}c}{{\pi}^2}}{\sqrt 
{\frac{M_{t_0}G^{\rm S}}{{\Xi}_0}}}
{\Big [} \int_{0}^{{\frac{{\xi}}{{\xi}_{\rm d}}}}+ 
\int_{{\frac{{\xi}}{{\xi}_{\rm d}}}}^{{\frac{{\Xi}_0}{{\xi}_{\rm d}}}}
{\Big ]}{\frac{x{\Big {\{}}e^{-x}{\rm Ei}(x)-e^x{\rm Ei}(-x)
{\Big {\}}}K{\Big (}2{\frac{{\sqrt{x{\xi}/{\xi}_{\rm d}}}}
{x+{\xi}/{\xi}_{\rm d}}}{\Big )}dx}{x+{\xi}/{\xi}_{\rm d}}}, 
\ena
for ${\xi}{\ll}{\Xi}_0$. Note that the magnitude of ${\Psi}({\xi})$ depends
critically on the upper limit of integration (if ${\xi}{\sim}{\Xi}_0$, a more 
accurate expression for ${\Psi}(\xi)$ may be found from equation (51)). Also
note that ${\Psi}(\xi)$ is non-singular everywhere, and that if 
${\bar {\Sigma}}_{\rm c}$ is small enough, $|{\Psi}(\xi)|$ may in principle 
dominate over $|{\Phi}(\xi)|$ for the whole disk. Finally, the full rotational 
curve may be found approximately from the formula (${\xi}{\ll}{\Xi}_0$)
\eqa
{\bar w}^2_{\rm circ}({\xi}){\approx}{\bar w}^2_{\rm real}({\xi})+
{\frac{c}{{\sqrt 2}{\pi}^2}}{\sqrt {\frac{M_{t_0}G^{\rm S}}{{\Xi}_0}}}{\Big [} 
\int_{0}^{{\frac{{\xi}}{{\xi}_{\rm d}}}}+ \int_{{\frac{{\xi}}{{\xi}_{\rm d}}}}
^{{\frac{{\Xi}_0}{{\xi}_{\rm d}}}}{\Big ]}x
{\Big {\{}}{\exp}(-x){\rm Ei}(x)-{\exp}(x){\rm Ei}(-x){\Big {\}}} \nonumber \\
{\times}{\Big [}{\frac{K{\Big (}2{\frac{{\sqrt{x{\xi}/{\xi}_{\rm d}}}}
{x+{\xi}/{\xi}_{\rm d}}}{\Big )}}{x+{\xi}/{\xi}_{\rm d}}}-
{\frac{E{\Big (}2{\frac{{\sqrt{x{\xi}/{\xi}_{\rm d}}}}
{x+{\xi}/{\xi}_{\rm d}}}{\Big )}}
{x-{\xi}/{\xi}_{\rm d}}}{\Big ]}dx.
\ena
Numerically, it is found that with increasing ${\xi}$, the integrals converge 
rather quickly towards a constant factor of $2{\pi}$, so for ${\xi}$ above 
about 8-10 disk lengths the associated induced contribution changes very 
little. This means that, for large distances, the expression for 
${\bar w}_{\rm circ}^2(\xi)$ does not depend significantly on the upper limit of 
integration, and that one automatically gets an asymptotically flat rotation
curve. Also notice that for small distances, the integrals yield a 
{\em negative} contribution to ${\bar w}_{\rm circ}^2(\xi)$.
\subsection{The correspondence with MOND}
The most basic feature of MOND is the postulation of a fundamental acceleration
scale $a_0$, observationally estimated to 
$a_0{\sim}(1-2){\times}10^{-10}$m/s$^2$. That is, for Keplerian accelerations 
well below $a_0$, the Newtonian acceleration $a_{\rm N}={\frac{G_{\rm N}M}{r^2}}$ 
in a spherically symmetric system is replaced by the MOND acceleration 
$a_{\rm M}={\frac{{\sqrt{G_{\rm N}Ma_0}}}{r}}$. With the help of results derived 
in the preceding section, we may now easily find a correspondence with MOND and 
actually {\em calculate} $a_0$ in the case of an exponential disk. From 
equation (65) we find that the asymptotic rotational speed ${\bar w}_{\infty}$
and the corresponding Keplerian acceleration are given by
\eqa
{\bar w}_{\infty}={\sqrt{{\frac{c}{\pi}}
{\sqrt {\frac{2M_{t_0}G^{\rm S}}{{\Xi}_0}}}}}, \qquad \Rightarrow \qquad
{\bar a}({\xi})={\frac{c}{\pi}}{\sqrt {\frac{2M_{t_0}G^{\rm S}}{{\Xi}_0}}}
{\frac{1}{\xi}}, \quad {\xi}{\gg}{\xi}_{\rm d}.
\ena
We are now able to compare equation (66) to the MOND acceleration. With 
$G^{\rm S}{\approx}G_{\rm N}$ measured at epoch $t_0$ we then get a 
correspondence if 
\eqa
a_0={\frac{2c^2}{{\pi}^2{\Xi}_0}}={\frac{2}{{\pi}^2}}cH_0{\approx}1.5{\times}
10^{-10}{\rm m/s}^2,
\ena
where the expression is valid for an exponential disk at epoch $t_0$ (with 
$H_0$ as the corresponding Hubble parameter), and where the final, numerical 
result is valid for the present epoch. (Note that $a_0$ is predicted to 
decrease with time, since the Hubble parameter decreases as the inverse of 
cosmic epoch.) This means that there is a correspondence between the model of 
a thin disk presented in this paper and MOND for the asymptotically flat part 
of the rotation curve. In particular, the fact that MOND successfully fits 
the observed baryonic Tully-Fisher relation [11] means that quasi-metric 
gravity does as well. That is, from equation (66) we get that
\eqa
{\bar w}_{\infty}^4={\frac{2M_{t_0}G^{\rm S}c^2}{{\pi}^2{\Xi}_0}}=
{\frac{2M_{t_0}G^{\rm S}}{{\pi}^2}}cH_0, \qquad \Rightarrow \qquad
M_{t_0}={\frac{{\pi}^2{\Xi}_0}{2G^{\rm S}c^2}}{\bar w}_{\infty}^4
={\frac{{\pi}^2}{2G^{\rm S}cH_0}}{\bar w}_{\infty}^4,
\ena
and since the constant of proportionality in the relation 
$M_{t_0}{\propto}{\bar w}^4_{\infty}$ is measured to be about 
$50$ M$_{\odot}$s$^4$/km$^4$ [11], it is straightforward to check that this 
result is in very good agreement with equation (68) (for the present epoch).
This justifies the definition of ${\bar {\sigma}}(-1)$ made in equation (35).
Note that, what enters into equation (68) is the active gravitational mass 
$M_t$ for material mass-energy (at epoch $t_0$); this increases linearly with 
cosmic epoch. Moreover, since the Hubble parameter decreases as the inverse of 
cosmic epoch, this means that ${\bar w}_{\infty}$ is constant; i.e., that it 
does not evolve with epoch. In other words, the quasi-metric model predicts 
that the sizes of metrically stationary galactic disks increase with epoch, 
but such that rotational speeds are unaffected.

Moreover, a FO at an arbitrary epoch $t_1$, using his locally measured values 
${\frac{t_1}{t_0}}G^{\rm S}$ and ${\frac{t_0}{t_1}}H_0$ of the gravitational 
``constant'' and the Hubble parameter, respectively, is predicted to measure 
the same slope of the local baryonic Tully-Fisher relation as will a FO at 
epoch $t_0$. This means that the slope of the baryonic Tully-Fisher relation is
predicted not to depend on epoch, and this seems to be confirmed by 
observations [12, 13]. (The analyses presented in these papers are based on the
standard cosmological framework, so some inconsistency with the QMF might be 
expected. But as long as theory-dependent gravitational physics (i.e., 
inconsistent with the QMF) is not used to infer masses, this should not 
matter too much.)
\section{The general weak-field, axisymmetric case}
Having analysed a flat disk, the question now is if the results of the previous
sections can be extended to the general, weak-field axisymmetric case (for
simplicity it is still assumed that the gravitational source contains a 
negligible amount of electromagnetic field energy). To answer that question, 
we will follow the standard method of breaking up an arbitrary axisymmetric 
matter distribution into a series of concentric, spherical shells (of 
negligible thickness) with corresponding surface densities. The total potential
at a given point will then be the sum of the potentials due to the collection 
of shells. Except for the limitation to axial symmetry, this procedure is the 
counterpart to the derivation of a general multipole expansion for the 
Newtonian case as given in [7], which we will follow closely.

The mathematical task is then to solve equation (10) (with $V=0$) interior 
respectively exterior to a given isolated shell, subject to suitable boundary
conditions. Since equation (10) becomes separable, we may write solutions in 
the form ${\bar B}({\rho},{\theta})=1+2c^{-2}{\Phi}({\rho},{\theta})$, were
$c^{-2}{\Phi}({\rho},{\theta}){\equiv}-{\bar F}({\rho}){\bar G}({\theta})$. 
General mode solutions ${\bar F}_{{\beta}{\pm}}({\rho})$, 
${\bar G}_{{\beta}{\pm}}({\theta})$ are given by equations (A.3), (A.4) in 
appendix A. We will select specific mode solutions ${\bar F}_n({\rho})$, 
${\bar G}_n({\theta})$, where $n=0,1,2,...$ are whole non-negative numbers, 
such that
\eqa
c^{-2}{\Phi}_n({\rho},{\theta})=-{\sqrt{{\frac{{\Xi}_0}{{\rho}}}}}{\Bigg [}
A_nP^{-n-{\frac{1}{2}}}_{\frac{1}{2}}{\Big (}
{\sqrt{1-{\frac{{\rho}^2}{{\Xi}_0^2}}}}{\Big )}+
B_nP^{n+{\frac{1}{2}}}_{\frac{1}{2}}{\Big (}
{\sqrt{1-{\frac{{\rho}^2}{{\Xi}_0^2}}}}{\Big )}{\Bigg ]}P_n({\cos}{\theta}),
\ena
where $A_n$ and $B_n$ are constants. We now construct solutions
${\Phi}_{\rm int}({\rho},{\theta})$, ${\Phi}_{\rm ext}({\rho},{\theta})$ 
respectively interior and exterior to a shell located at 
${\rho}={\rho}_{\rm s}$, by summing up suitable mode solutions. That is, by 
requiring that the interior solution should be regular at the centre of the 
shell, we find that
\eqa
c^{-2}{\Phi}_{\rm int}({\rho},{\theta})=-{\sqrt{{\frac{{\Xi}_0}{{\rho}}}}}
\sum_{n=0}^{\infty}A_nP^{-n-{\frac{1}{2}}}_{\frac{1}{2}}{\Big (}
{\sqrt{1-{\frac{{\rho}^2}{{\Xi}_0^2}}}}{\Big )}P_n({\cos}{\theta}), \qquad
{\rho}<{\rho}_{\rm s}.
\ena
The exterior solution is found by requiring that 
${\Phi}_{\rm ext}({\Xi}_0,{\theta})$ must vanish. So, for ${\rho}>{\rho}_{\rm s}$, 
\eqa
c^{-2}{\Phi}_{\rm ext}({\rho},{\theta})=-{\sqrt{{\frac{{\Xi}_0}{{\rho}}}}}
\sum_{n=0}^{\infty}{\Bigg [}D_nP^{-n-{\frac{1}{2}}}_{\frac{1}{2}}{\Big (}
{\sqrt{1-{\frac{{\rho}^2}{{\Xi}_0^2}}}}{\Big )}+
B_nP^{n+{\frac{1}{2}}}_{\frac{1}{2}}{\Big (}
{\sqrt{1-{\frac{{\rho}^2}{{\Xi}_0^2}}}}{\Big )}{\Bigg ]}P_n({\cos}{\theta}),
\ena
where the constants $D_n$ are given by the expression $D_n=-B_n
P^{n+{\frac{1}{2}}}_{\frac{1}{2}}(0)/P^{-n-{\frac{1}{2}}}_{\frac{1}{2}}(0)$, which vanishes 
for all even $n$ (see equation (A.5)).

Now the surface density ${\bar {\Sigma}}_{\rm s}({\rho}_{\rm s},{\theta})$ of the 
shell can be expressed as a sum of modes; this yields counterparts to 
equations (23)-(25) valid for a thin disk. We thus have
\eqa
{\bar {\Sigma}}_{\rm s}({\rho}_{\rm s},{\theta})={\frac{1}{2}} 
\sum_{n=0}^{\infty}(2n+1){\bar {\Sigma}}_{{\rm s}n}({\rho}_{\rm s})
P_n({\cos}{\theta}), \quad {\bar {\Sigma}}_{{\rm s}n}({\rho}_{\rm s})= 
\int_0^{\pi}{\bar {\Sigma}}_{\rm s}({\rho}_{\rm s},{\theta}')
P_n({\cos}{\theta}'){\sin}{\theta}'d{\theta}'.
\ena
To determine the constants $A_n$ and $B_n$, we now apply Gauss' theorem
across the shell. Assuming a weak gravitational field (${\bar B}{\approx}1$)
we then have
\eqa
{\sqrt{1-{\frac{{\rho}_{\rm s}^2}{{\Xi}_0^2}}}}{\Bigg [}
{\Big (}{\frac{{\partial}{\Phi}_{\rm ext}}{{\partial}{\rho}}}{\Big )}_
{{\rho}={\rho}_{\rm s}}-{\Big (}
{\frac{{\partial}{\Phi}_{\rm int}}{{\partial}{\rho}}}{\Big )}_
{{\rho}={\rho}_{\rm s}}{\Bigg ]}=4{\pi}G^{\rm S}{\bar {\Sigma}}_{\rm s}
({\rho}_{\rm s},{\theta}).
\ena
Since the potential must be continuous over the shell, we have
${\Phi}_{\rm ext}({\rho}_{\rm s},{\theta})={\Phi}_{\rm int}({\rho}_{\rm s},{\theta})$.
Furthermore, due to the orthogonality of the Legendre polynomials, this means 
that
\eqa
{\Big (}A_n+{\frac{P^{n+{\frac{1}{2}}}_{\frac{1}{2}}(0)}
{P^{-n-{\frac{1}{2}}}_{\frac{1}{2}}(0)}}B_n{\Big )}
P^{-n-{\frac{1}{2}}}_{\frac{1}{2}}{\Big (}
{\sqrt{1-{\frac{{\rho}_{\rm s}^2}{{\Xi}_0^2}}}}{\Big )}
=B_nP^{n+{\frac{1}{2}}}_{\frac{1}{2}}{\Big (}
{\sqrt{1-{\frac{{\rho}_{\rm s}^2}{{\Xi}_0^2}}}}{\Big )}.
\ena
A second equation relating the constants $A_n$ and $B_n$ can be found from
equation (73) by inserting equation (72). Calculating the derivatives and
using recurrence formulae valid for Legendre functions, one finds that
(since the Legendre polynomials are orthogonal)
\eqa
(1-n^2)P^{n-{\frac{1}{2}}}_{\frac{1}{2}}{\Big (}
{\sqrt{1-{\frac{{\rho}_{\rm s}^2}{{\Xi}_0^2}}}}{\Big )}B_n&+&
{\Big (}A_n+{\frac{P^{n+{\frac{1}{2}}}_{\frac{1}{2}}(0)}
{P^{-n-{\frac{1}{2}}}_{\frac{1}{2}}(0)}}
B_n{\Big )}P^{-n+{\frac{1}{2}}}_{\frac{1}{2}}{\Big (}
{\sqrt{1-{\frac{{\rho}_{\rm s}^2}{{\Xi}_0^2}}}}{\Big )} \nonumber \\
&&=(2n+1){\sqrt{{\frac{{\rho}_{\rm s}}{{\Xi}_0}}}}
{\frac{{\bar {\Sigma}}_{{\rm s}n}({\rho}_{\rm s})}{{\bar {\Sigma}}_*}}.
\ena
One may find explicit expressions for $A_n$ and $B_n$ from equations (74)
and (75). We get
\eqa
A_n=(2n+1){\sqrt{\frac{{\rho}_{\rm s}}{{\Xi}_0}}}
{\frac{{\bar {\Sigma}}_{{\rm s}n}({\rho}_{\rm s})}{{\bar {\Sigma}}_*}}
g_n({\rho}_{\rm s}), \qquad
B_n=(2n+1){\sqrt{\frac{{\rho}_{\rm s}}{{\Xi}_0}}}
{\frac{{\bar {\Sigma}}_{{\rm s}n}({\rho}_{\rm s})}{{\bar {\Sigma}}_*}}
f_n({\rho}_{\rm s}),
\ena
\eqa
f_n({\rho}_{\rm s}){\equiv}{\frac{P^{-n-{\frac{1}{2}}}_{\frac{1}{2}}{\big (}
y_{\rm s}{\big )}}{(1-n^2)P^{-n-{\frac{1}{2}}}_{\frac{1}{2}}{\big (}
y_{\rm s}{\big )}P^{n-{\frac{1}{2}}}_{\frac{1}{2}}{\big (}y_{\rm s}{\big )}+
P^{n+{\frac{1}{2}}}_{\frac{1}{2}}{\big (}y_{\rm s}{\big )}
P^{-n+{\frac{1}{2}}}_{\frac{1}{2}}{\big (}y_{\rm s}{\big )}}}, \qquad 
y_{\rm s}{\equiv}{\sqrt{1-{\frac{{\rho}_{\rm s}^2}{{\Xi}_0^2}}}},
\ena
\eqa
g_n({\rho}_{\rm s}){\equiv}{\Bigg [}
{\frac{P^{n+{\frac{1}{2}}}_{\frac{1}{2}}{\big (}
y_{\rm s}{\big )}}{P^{-n-{\frac{1}{2}}}_{\frac{1}{2}}{\big (}
y_{\rm s}{\big )}}}-{\frac{P^{n+{\frac{1}{2}}}_{\frac{1}{2}}{\big (}
0{\big )}}{P^{-n-{\frac{1}{2}}}_{\frac{1}{2}}{\big (}
0{\big )}}}{\Bigg ]}f_n({\rho}_{\rm s}).
\ena
Equations (76)-(78) may now be inserted into equations (70), (71) to get
complete expressions for ${\Phi}_{\rm int}({\rho},{\theta})$ and
${\Phi}_{\rm ext}({\rho},{\theta})$. This yields the potential generated by a
single, infinitesimally thin shell.

To evaluate the potential generated by an entire collection of shells filling
space, we let ${\delta}{\bar {\Sigma}}_{\rm s}({\rho}_{\rm s},{\theta})$ and
${\delta}{\bar {\Sigma}}_{{\rm s}n}({\rho}_{\rm s})$ denote the relevant quantities
for a shell lying between ${\rho}_{\rm s}$ and
${\rho}_{\rm s}+{\delta}{\rho}_{\rm s}$. From equation (72) we then have
(inserting ${\delta}{\bar {\Sigma}}_{\rm s}({\rho}_{\rm s},{\theta}')=
{\bar {\varrho}}_{\rm m}({\rho}_{\rm s},{\theta}'){\delta}{\rho}_{\rm s}/
{\sqrt{1-{\frac{{\rho}_{\rm s}^2}{{\Xi}_0^2}}}}$)
\eqa
{\delta}{\bar {\Sigma}}_{{\rm s}n}({\rho}_{\rm s})= 
\int_0^{\pi}{\bar {\varrho}}_{\rm m}({\rho}_{\rm s},{\theta}')
P_n({\cos}{\theta}'){\sin}{\theta}'d{\theta}'{\frac{{\delta}{\rho}_{\rm s}}
{{\sqrt{1-{\frac{{\rho}_{\rm s}^2}{{\Xi}_0^2}}}}}}{\equiv}
{\frac{2{\bar {\varrho}}_{{\rm m}n}({\rho}_{\rm s}){\delta}{\rho}_{\rm s}}
{{\sqrt{1-{\frac{{\rho}_{\rm s}^2}{{\Xi}_0^2}}}}}},
\ena
where ${\bar {\varrho}}_{{\rm m}}({\rho}_{\rm s},{\theta})$ is the properly 
scaled density of (active) mass [4, 5]. Substituting equation (79) into the
complete expressions for the corresponding potentials 
${\delta}{\Phi}_{\rm int}({\rho},{\theta})$ and
${\delta}{\Phi}_{\rm ext}({\rho},{\theta})$ and integrating over ${\rho}_{\rm s}$,
we find that (with $x_{\rm s}{\equiv}{\frac{{\rho}_{\rm s}}{{\Xi}_0}}$,
$y{\equiv}{\sqrt{1-{\frac{{\rho}^2}{{\Xi}_0^2}}}}$ and
${\bar {\varrho}}_{{\rm m }*}{\equiv}{\frac{c^2}{4{\pi}G^{\rm S}{\Xi}_0^2}}$)
\eqa
c^{-2}{\Phi}({\rho},{\theta})=c^{-2} \sum_{{\rho}_{\rm s}=0}^{\rho}
{\delta}{\Phi}_{\rm ext}({\rho},{\theta})+c^{-2} \sum_{{\rho}_{\rm s}=
{\rho}}^{{\Xi}_0}{\delta}{\Phi}_{\rm int}({\rho},{\theta}) \nonumber \\
=-{\sqrt{{\frac{{\Xi}_0}{{\rho}}}}}\sum_{n=0}^{\infty}
{\frac{(2n+1)}{{\bar {\varrho}}_{{\rm m}*}}}P_n({\cos}{\theta}){\Bigg [}
{\Big \{}P^{n+{\frac{1}{2}}}_{\frac{1}{2}}{\big (}y{\big )}-
{\frac{P^{n+{\frac{1}{2}}}_{\frac{1}{2}}{\big (}0{\big )}}
{P^{-n-{\frac{1}{2}}}_{\frac{1}{2}}{\big (}0{\big )}}}
P^{-n-{\frac{1}{2}}}_{\frac{1}{2}}{\big (}y{\big )}{\Big \}} \nonumber \\
{\times} \int_0^{{\frac{\rho}{{\Xi}_0}}}
{\frac{{\sqrt{x_{\rm s}}}f_n(x_{\rm s}){\bar {\varrho}}_{{\rm m}n}(x_{\rm s})
dx_{\rm s}}{\sqrt{1-x_{\rm s}^2}}}+
P^{-n-{\frac{1}{2}}}_{\frac{1}{2}}{\big (}y{\big )}
\int_{{\frac{\rho}{{\Xi}_0}}}^1
{\frac{{\sqrt{x_{\rm s}}}g_n(x_{\rm s}){\bar {\varrho}}_{{\rm m}n}(x_{\rm s})
dx_{\rm s}}{\sqrt{1-x_{\rm s}^2}}}{\Bigg ]}.
\ena
Equation (80) is the axisymmetric counterpart to the similar general multipole 
expansion formula in Newtonian theory; this is given explicitly in [7].
These two expressions have a correspondence in the limit
${\Xi}_0{\rightarrow}{\infty}$. For example, the potential at ${\rho}=0$ 
is given by $c^{-2}{\Phi}(0)=-{\frac{1}{{\bar {\varrho}}_{{\rm m }*}}} 
\int_0^1x_{\rm s}{\bar {\varrho}}_{{\rm m }0}(x_{\rm s})dx_{\rm s}$, the same as the
Newtonian expression in said limit. This means that as expected, equation (80)
can not describe DM-effects.

Moreover, due to the boundary condition at ${\rho}={\Xi}_0$, there is no 
obvious way to define an associated (volume) density as a counterpart to 
equation (35). But if no associated density can be defined, one cannot motivate
a definition similar to equation (46) for the associated induced (volume)
density. So it would seem that, the explanation of DM-effects presented in this
paper should exclusively be connected to disks or other structures that do not 
satisfy the boundary condition ${\Phi}({\Xi}_0)=0$, answering the question 
asked at the beginning of this section.

However, DM-effects are seen in galaxies other than spiral galaxies. In 
particular, observations indicate that dwarf spheroidal galaxies are the most 
DM-dominated systems ever found [14]. Yet the general observational status for 
the existence of DM in elliptical galaxies is more complicated than for spiral 
galaxies, since some ordinary elliptical galaxies apparently lack significant 
amounts of it [15], while others seem to have plenty [16]. It still remains the
challenging task of explaining these observations without DM. 
\section{Gravitational lensing}
For a sufficiently weak gravitational field, rotation curves can be calculated 
accurately enough using the auxiliary metric family 
${\bf {\bar g}}_t$ rather than the full ``physical'' metric family ${\bf g}_t$.
However, when calculating deflection of light, or gravitational lensing, it
is not sufficient to know ${\bf {\bar g}}_t$, even if the gravitational field
is weak. But to calculate gravitational lensing, it is fortunately not 
necessary to know ${\bf g}_t$ in full; as we shall see, a suitable 
approximation will be sufficient.

The general formulae describing the transformation ${\bf {\bar g}}_t
{\rightarrow}{\bf g}_t$ are given by [2, 3]
\eqa
g_{(t)00}&=&{\Big (}1-{\frac{v^2}{c^2}}{\Big )}^2{\bar g}_{(t)00}, \\
g_{(t)0j}&=&{\Big (}1-{\frac{v^2}{c^2}}{\Big )}{\Big [}{\bar g}_{(t)0j}
+{\frac{t}{t_0}}{\frac{2{\frac{v}{c}}}{1-{\frac{v}{c}}}}({\bar e}^i_{\cal F}
{\bar N}_{(t)i}){\bar \omega}_{{\cal F}j}{\Big ]}, \\
g_{(t)ij}&=&{\bar g}_{(t)ij}+{\frac{t^2}{t_0^2}}{\frac{4{\frac{v}{c}}}
{(1-{\frac{v}{c}})^2}}{\bar \omega}_{{\cal F}i}{\bar \omega}_{{\cal F}j},
\ena
where ${\bf {\bar e}}_{\cal F}{\equiv}
{\frac{t_0}{t}}{\bar e}^i_{\cal F}{\frac{\partial}{{\partial}x^i}}$ and 
${\bf {\bar \omega}}_{\cal F}{\equiv}{\frac{t}{t_0}}{\bar \omega}_{{\cal F}i}dx^i$
are unit vector and covector fields, respectively, along the 3-vector field
${\bf {\bar x}}_{\cal F}$ found from the set of linear algebraic equations
\eqa
{\Big [}{\bar a}_{{\cal F}{\mid}k}^k+c^{-2}{\bar a}_{{\cal F}k}
{\bar a}_{\cal F}^k{\Big ]}{\bar x}_{\cal F}^j-
{\Big [}{\bar a}_{{\cal F}{\mid}k}^j+c^{-2}{\bar a}_{{\cal F}k}
{\bar a}_{\cal F}^j{\Big ]}{\bar x}_{\cal F}^k-2{\bar a}_{\cal F}^j=0,
\ena
and where $v{\equiv}c^{-1}{\sqrt{{\bar a}_{{\cal F}k}{\bar a}_{\cal F}^k
{\bar x}_{{\cal F}i}{\bar x}_{\cal F}^i}}$. For a weak gravitational field and for 
distances much smaller than ${\Xi}_0$ (at epoch $t_0$), we have that $v{\ll}c$,
so to a good approximation we may neglect terms of order 2 or higher in the 
small quantity $v/c$. This means that we may set (assuming 
${\bar N}_{(t)i}{\approx}0$)
\eqa
g_{(t)00}{\approx}{\bar g}_{(t)00}, \qquad g_{(t)0j}{\approx}0, \qquad
g_{(t)ij}{\approx}{\bar g}_{(t)ij}+4{\frac{t^2}{t_0^2}}
{\frac{v}{c}}{\bar \omega}_{{\cal F}i}{\bar \omega}_{{\cal F}j}.
\ena
To get explicit formulae, it is easiest to solve the set of equations (84) 
using standard methods, in spherical coordinates. One may then transform to 
cylindrical coordinates. If one additionally assumes a weak field in vacuum, 
and neglects all terms proportional to ${\Xi}_0^{-2}$, one may give 
${\bar x}_{{\cal F}}^{\xi}$ and ${\bar x}_{{\cal F}}^z$ as series expansions in $z$.
Terminating the series after the linear term, yields (using the fact that in 
vacuum, ${\bar B},_{zz}{\approx}-{\bar B},_{{\xi}{\xi}}-
{\frac{1}{\xi}}{\bar B},_{\xi}$ from equation (13))
\eqa
{\bar x}_{{\cal F}}^{\xi}&{\approx}& 
{\frac{-2{\Big [}({\bar B},_{{\xi}{\xi}}+{\frac{1}{\xi}}{\bar B},_{\xi})
{\bar B},_{\xi}+{\bar B},_{{\xi}z}{\bar B},_z{\Big ]}}{{\Big (}
{\bar B},_{{\xi}{\xi}}+{\frac{1}{\xi}}{\bar B},_{\xi}{\Big )}{\bar B},_{{\xi}
{\xi}}+({\bar B},_{{\xi}z})^2}}+{\frac{z}{\xi}}{\bar x}_{\cal F}^z + O(z^2), \\
{\bar x}_{{\cal F}}^z&{\approx}&
{\frac{2{\Big [}{\bar B},_{{\xi}{\xi}}{\bar B},_z
-{\bar B},_{{\xi}z}{\bar B},_{\xi}{\Big ]}}{
{\Big (}{\bar B},_{{\xi}{\xi}}+{\frac{1}{\xi}}{\bar B},_{\xi}{\Big )}
{\bar B},_{{\xi}{\xi}}+({\bar B},_{{\xi}z})^2}} + O(z^2),
\ena
where the derivatives may be calculated approximately to first order in $|z|$
from the expression
\eqa
{\bar B}({\xi},z)&{\approx}&1+{\frac{2}{c^2}}{\Big (}{\Phi}({\xi})+
{\Psi}({\xi}){\Big )}+{\frac{4{\pi}G^{\rm S}}{c^2}}{\Big (}
{\bar {\Sigma}}({\xi})+{\bar {\sigma}}^{\rm i}({\xi}){\Big )}|z| \nonumber \\
&-&{\frac{1}{c^2}}{\Big [}{\Phi},_{{\xi}{\xi}}+{\frac{1}{\xi}}{\Phi},_{{\xi}}
+{\Psi},_{{\xi}{\xi}}+{\frac{1}{\xi}}{\Psi},_{{\xi}}{\Big ]}z^2
+O(|z|^3),
\ena
obtained from equation (54) and the approximation for ${\bar B},_{zz}$ shown
above. Furthermore, since the quantities ${\bar \omega}_{{\cal F}{\xi}}$ and 
${\bar \omega}_{{\cal F}z}$ may be found to the relevant accuracy from the 
definition
\eqa
{\bar \omega}_{{\cal F}j}={\bar h}_{(t_0)ij}{\bar e}^i_{\cal F}{\equiv}
{\bar h}_{(t_0)ij}{\frac{{\bar x}_{\cal F}^i}{{\mid}{\bf {\bar x}}_{\cal F}
{\mid}}}{\equiv}{\frac{{\bar x}_{{\cal F}j}}{{\mid}{\bf {\bar x}}_{\cal F}
{\mid}}}, \qquad {\mid}{\bf {\bar x}}_{\cal F}{\mid}{\equiv}
{\sqrt{{\bar x}_{{\cal F}k}{\bar x}_{\cal F}^k}},
\ena
we may also calculate the relevant approximation of the desired quantities 
$g_{(t)ij}$ from equation (85), i.e.,
\eqa
g_{(t)ij}{\approx}{\bar g}_{(t)ij}+4{\frac{t^2}{t_0^2}}
{\frac{v}{c}}{\bar \omega}_{{\cal F}i}{\bar \omega}_{{\cal F}j}{\approx}
{\bar g}_{(t)ij}+4{\frac{t^2}{t_0^2}}{\sqrt{
{\frac{{\bar B},_k{\bar B},^k}{{\bar x}_{{\cal F}s}{\bar x}_{\cal F}^s}}}}
{\bar x}_{{\cal F}i}{\bar x}_{{\cal F}j}.
\ena
From the geodesic equation, we may now calculate the gravitational bending of a
light ray grazing the plane of the disk, since in this case, $|z|$ is small 
enough so that the approximation given in equation (88) is valid. To deal with 
the opposite situation, where the light path is nearly orthogonal to the disk 
plane, the approximations given in equations (86)-(88) may not be sufficient;
then one must include terms of higher order in $|z|$.

Finally, we notice that in GR, the weak field form of the metric outside the 
disk is assumed to take the form
\eqa
ds^2=-(1+{\frac{2}{c^2}}{\Phi}_{\rm N})(dx^0)^2+
(1-{\frac{2}{c^2}}{\Phi}_{\rm N})(dx^2+dy^2+dz^2),
\ena
where ${\Phi}_{\rm N}$ is the Newtonian potential for the sum of visible and DM.
But in general, the quantities $g_{(t_0)ij}$ as found from equation (90) will
not correspond to their counterparts given by equation (91). This means that 
any observationally based mapping of DM distributions using gravitational 
lensing, assuming the weak field approximation obtained from GR, is explicitly 
theory-dependent and may give misleading results.
\section{Discussion}
For many years, it has been known that galactic dynamics is incompatible with
a straightforward application of Newtonian theory to visible matter. However, 
the most glaring discrepancies between observations and theory can be removed 
by assuming the existence of galactic DM. Since the introduction of DM can be 
done without making radical changes to the standard theoretical framework
underlying mainstream astrophysics, this is currently the preferred approach.
On the other hand, MOND interpreted as an empirical recipe, has an 
impressive successful record when predicting galactic phenomena. But the 
connection between MOND and fundamental physics has been unclear so far.

Contrary to other approaches, the explanation of some galactic phenomena given 
in this paper has not assumed any empirical aspect of galactic dynamics as 
input to the model. Rather, while formally belonging to the 
``modified gravity'' category, the model comes directly from the weak field 
approximation of the QMF, without any extra modifications of the theory. (The 
only extra assumption made, is that the associated induced surface density 
${\bar{\sigma}}^{\rm i}(u)$ should be treated as a gravitating source in the 
field equations.) The main reason why this is possible is that according to the
QMF, the Universe is finite and ``small'', so that boundary conditions depend 
crucially on the shape of the matter distribution. (A finite and ``small'' 
Universe is incompatible with cosmological data as interpreted within the 
standard framework, therefore the DM explanation given here is compatible with 
the weak field limit of the QMF but not with standard cosmology.) In 
particular, for a flat disk we found that ${\Phi}(u=-1){\neq}0$ in equation 
(32), defining a specific velocity scale dependent on the total mass of the 
disk. To be able to define the associated surface density ${\bar{\sigma}}(u)$ 
in equation (35), it is essential that this velocity scale does not vanish. 
However, by construction, it does vanish for the general axisymmetric matter 
distribution considered in section 5, so flat disks seem to be an exceptional 
case. Anyway, that case should apply to all thin disks (even if they are not 
exactly flat).

The other crucial feature is the existence of an induced matter surface density
${\bar{\Sigma}}^{\rm i}(u)$ directly dependent on the real matter surface 
density as shown in equation (45). This, together with the existence of 
${\bar{\sigma}}(u)$, is sufficient to define the associated induced surface 
density ${\bar{\sigma}}^{\rm i}(u)$ playing the role of DM. The introduction of 
${\bar{\sigma}}^{\rm i}(u)$ may seem ``contrived'' to some people. However, the 
facts that the associated induced surface density corresponding to an 
exponential disk automatically yields an asymptotically flat rotation curve 
and a correspondence with MOND are {\em calculated} results, and not put in by 
hand {\em a priori}. It was not at all obvious that these results would be 
possible.

It has been claimed [17] that observations of gravitational lensing in the
colliding clusters 1E0657-56 (the Bullet cluster) represent a ``direct'' 
detection of DM, since these observations indicate that the DM is associated 
with the regions containing the field galaxies rather than with the regions 
containing the more massive gas making up the bulk of the cluster. However, as 
we have seen, in the QMF, the existence of any sort of ``phantom'' matter 
density similar to ${\bar{\sigma}}^{\rm i}(u)$, playing the role of DM, is 
crucially dependent on the {\em shape} of the matter distribution. This means 
that, e.g., a large nearly spherical or spheroidal mass distribution of gas, 
should not necessarily be associated with much DM as inferred from 
gravitational lensing. So, since the colliding gas clouds in the Bullet cluster
shown in [17] do not seem to have shapes that could in any way resemble disks, 
this might be a natural explanation of why they do not seem to be associated 
with much DM. But of course, further justification of this explanation will be 
necessary, together with an explanation of why dwarf spheroidals and some 
elliptic galaxies seem to be DM-dominated. Anyway, the mere existence of such 
an explanation shows that the interpretation of the Bullet cluster observations
is not theory-independent, so citing them as definite evidence of the existence
of DM is unjustified. 

In light of the results found in this paper, it seems that some lines of 
argument favouring DM over modified gravity have been shown to be invalid. 
First, there now exists a natural correspondence between MOND and fundamental 
physics. This indicates that at least some galactic phenomenology has its basis
in {\em geometry} rather than in the properties of some unknown exotic 
particle. Second, while the correspondence with MOND works for spiral galaxies,
this does not imply that such a correspondence is necessarily valid for other 
types of galaxies or galaxy clusters. This means that it may be possible to 
share MOND's successes but not necessarily its failures. (Further work should 
be done to see if this is indeed the case.) Third, some specific observations 
(of, e.g., the Bullet cluster) have been hailed as the ultimate ``direct 
proof'' that DM prevails over modified gravity. But since standard 
interpretations of such observations are model-dependent, this is not true.

There is now near consensus that galactic phenomenology has it basis in some 
suitable nonbaryonic particle, despite the fact that no such particle has ever 
been detected directly. The fact that one needs a DM-component in the early
Universe, in order to have standard cosmology agree with primordial
nucleosynthesis and a standard analysis of the cosmic microwave background, 
has reinforced this consensus. That is, many astrophysicists are of the opinion 
that indirect observations are sufficient to rule out any alternative 
explanations. This means that for many astrophysicists, taking the focus off 
DM and recognising the merits of modified gravity, is a radical move that is 
out of the question, regardless of how  contrived and {\em ad hoc} explanations
based on DM might be. This is a perilous attitude, since interpretations of 
indirect observations are often crucially theory-dependent. Another example of 
this is interpretations of some indirect observations made in the solar system 
that are also presented as indisputable facts, even if alternative 
interpretations are not ruled out [4]. So it would seem that, for at least 
some parts of astrophysics, there is a serious lack of critical assessment of 
basic assumptions underlying mainstream knowledge. One may hope that this will 
change in the future.
\\ [4mm]
{\bf References} \\ [1mm]
{\bf [1]} A. Aguirre, C.P. Burgess, A. Friedland, and  D. Nolte, \\
{\hspace*{6.3mm}}{\em Class. Quantum Grav.} {\bf 18}, R223 (2001). \\
{\bf [2]} D. {\O}stvang, {\em Grav. \& Cosmol.} {\bf 11}, 205 (2005) 
(gr-qc/0112025). \\
{\bf [3]} D. {\O}stvang, {\em Doctoral thesis}, NTNU, 2001 (gr-qc/0111110). \\
{\bf [4]} D. {\O}stvang, {\em Grav. \& Cosmol.} {\bf 13}, 1 (2007) 
(gr-qc/0201097). \\
{\bf [5]} D. {\O}stvang, 
{\em Acta Physica Polonica B} {\bf 39}, 1849 (2008) (gr-qc/0510085). \\
{\bf [6]} A. Toomre, {\em Astrophys. Journ.} {\bf 138}, 385 (1963).  \\
{\bf [7]} J. Binney and S. Tremaine, {\em Galactic dynamics},
Princeton University Press (1987). \\
{\bf [8]} M. Abramowitz and I.A. Stegun, {\em Handbook of mathematical
functions}, Dover (1972). \\
{\bf [9]} A.J. Jerri, {\em Integral and discrete transformations with 
applications and error analysis}, \\ 
{\hspace*{6.4mm}}Marcel Dekker Inc. (1992). \\
{\bf [10]} G.N. Watson, {\em J. London Math. Soc.} {\bf 8}, 289 (1933). \\
{\bf [11]} S.S. McGaugh, {\em Astrophys. Journ.} {\bf 632}, 859 (2005). \\
{\bf [12]} H. Flores {\em et al.}, {\em A{\&}A} {\bf 455}, 107 (2006)
(astro-ph/0603563). \\
{\bf [13]} M. Puech {\em et al.}, {\em A{\&}A} {\bf 510}, A68 (2010)
(arXiv:0903.3961). \\
{\bf [14]} S.S. McGaugh and J. Wolf, {\em Astrophys. Journ.} {\bf 722}, 248 
(2010) (arXiv:1003.3448). \\
{\bf [15]} A. Romanowsky {\em et al.}, {\em Science} {\bf 301}, 1696 (2003). \\
{\bf [16]} T. Bridges {\em et al.}, {\em MNRAS} {\bf 373}, 157 (2006). \\
{\bf [17]} D. Clowe {\em et al.}, {\em Astrophys. Journ.} {\bf 648}, L109 
(2006) (astro-ph/0608407).
\appendix
\renewcommand{\theequation}{\thesection.\arabic{equation}}
\section{Boundary behaviour of mode solutions}
\setcounter{equation}{0}
In this Appendix, we list mode solutions of the vacuum field equations for
axially symmetric, metrically static, isolated sources with various ``pure'' 
values of their multipole moments. As we shall see, these solutions can be 
classified into two groups; those that admit the boundary condition 
${\bar B}({\Xi}_0)=1$ and those that do not. 

Starting with equation (10), we may set ${\bar V}=0$ for a non-rotating source 
(this is a good approximation for slowly rotating sources and weak 
gravitational fields also). Equation (10) then becomes separable, i.e., 
solutions of it can be written in the form ${\bar B}({\rho},{\theta})=1-
2{\bar F}({\rho}){\bar G}({\theta})$. The new functions ${\bar F}$ and 
${\bar G}$ must then satisfy the ordinary differential equations
\eqa
(1-{\frac{{\rho}^2}{{\Xi}_0^2}}){\frac{d^2{\bar F}}{d{\rho}^2}}+
{\frac{2}{\rho}}(1-{\frac{3{\rho}^2}{2{\Xi}_0^2}}){\frac{d{\bar F}}{d{\rho}}}
-{\frac{{\beta}}{{\rho}^2}}{\bar F}=0,
\ena
\eqa
{\frac{d^2{\bar G}}{d{\theta}^2}}+{\cot}{\theta}{\frac{d{\bar G}}{d{\theta}}}
+{\beta}{\bar G}=0,
\ena
where ${\beta}$ is some complex-valued constant. Restricting ${\beta}$ to be
real and requiring that ${\beta}{\geq}-{\frac{1}{4}}$, the general solutions 
of equations (A.1) and (A.2) may be written in the form
\eqa
{\bar F}_{{\beta}{\pm}}({\rho})={\sqrt{{\frac{{\Xi}_0}{{\rho}}}}}{\Big [}
c_{{\beta}_{\pm}}P^{{\pm}{\sqrt{{\beta}+{\frac{1}{4}}}}}_
{\frac{1}{2}}{\Big (}{\sqrt{1-{\frac{{\rho}^2}{{\Xi}_0^2}}}}{\Big )}
+c_{{\beta}_{\pm}}^{\rm i}Q^{{\pm}{\sqrt{{\beta}+{\frac{1}{4}}}}}_{\frac{1}{2}}
{\Big (}{\sqrt{1-{\frac{{\rho}^2}{{\Xi}_0^2}}}}{\Big )}{\Big ]},
\ena
\eqa
{\bar G}_{{\beta}{\pm}}({\theta})=C_{{\beta}_{\pm}}
P_{{\pm}{\sqrt{{\beta}+{\frac{1}{4}}}}-
{\frac{1}{2}}}({\cos}{\theta})+C_{{\beta}_{\pm}}^{\rm i}Q_{{\pm}{\sqrt{{\beta}+
{\frac{1}{4}}}}-{\frac{1}{2}}}({\cos}{\theta}),
\ena
where $P_{\nu}^{\mu}(x)$, $Q_{\nu}^{\mu}(x)$ are the usual associated Legendre
functions of the first and second kind, respectively, and where 
$c_{{\beta}_{\pm}}$, $c_{{\beta}_{\pm}}^{\rm i}$, $C_{{\beta}_{\pm}}$ and 
$C_{{\beta}_{\pm}}^{\rm i}$ are (dimensionless) constants. Note that the solutions 
(A.3) and (A.4) are real-valued functions, and so are 
$P_{\nu}^{\mu}(x)$, $Q_{\nu}^{\mu}(x)$ (for real ${\mu}$, ${\nu}$) since they are 
defined on the cut $(-1,1)$ by averaging the relevant limiting values of the 
corresponding complex-valued quantities $P_{\nu}^{\mu}(z)$, $Q_{\nu}^{\mu}(z)$ [8].

We will now find solutions (A.3) admitting the boundary condition 
${\bar F}_{{\beta}{\pm}}({\Xi}_0)=0$. To do that, we first notice that, since 
$P^{\frac{1}{2}}_{\frac{1}{2}}({\sqrt{1-{\frac{{\rho}^2}{{\Xi}_0^2}}}})=
{\sqrt{{\frac{2{\Xi}_0(1-{\frac{{\rho}^2}{{\Xi}_0^2}})}{{\pi}{\rho}}}}}$ and
since only the trivial constant solution is obtained from
$P^{-{\frac{1}{2}}}_{\frac{1}{2}}({\sqrt{1-{\frac{{\rho}^2}{{\Xi}_0^2}}}})=
{\sqrt{{\frac{2{\rho}}{{\pi}{\Xi}_0}}}}$, choosing the $+$-sign,
with $C_{0_+}=1$ and a suitable choice of $c_{0_+}$, the value ${\beta}=0$ 
corresponds to the unique spherically symmetric solution found in ref. [4]. 
That solution is unique since the function $Q^{-{\frac{1}{2}}}_{\frac{1}{2}}(x)$ 
differs from $P^{\frac{1}{2}}_{\frac{1}{2}}(x)$
only by a numerical factor, and since $Q^{\frac{1}{2}}_{\frac{1}{2}}
({\sqrt{1-{\frac{{\rho}^2}{{\Xi}_0^2}}}})$ 
again only yields the trivial constant solution.  
Besides, since the function $Q_0({\cos}{\theta})$ has 
singularities whenever ${\theta}=0$ or ${\theta}={\pi}$, the corresponding 
solution cannot be physical (but might be considered as a mode solution). 
Moreover, $P_{-1}({\cos}{\theta})=P_{0}({\cos}{\theta})$ gives nothing new, and 
since $Q_{-1}({\cos}{\theta})$ is undefined, that expression cannot be 
considered at all.

Secondly we notice that, since choosing ${\beta}=2$ yields 
$P^{\frac{3}{2}}_{\frac{1}{2}}({\sqrt{1-{\frac{{\rho}^2}{{\Xi}_0^2}}}})=
-{\sqrt{{\frac{2{\Xi}^3_0}{{\pi}{\rho}^3}}}}$ and $P^{-{\frac{3}{2}}}_{\frac{1}{2}}
({\sqrt{1-{\frac{{\rho}^2}{{\Xi}_0^2}}}})={\sqrt{{\frac{{\Xi}^3_0}
{2{\pi}{\rho}^3}}}}{\Big (}{\arccos}({\sqrt{1-{\frac{{\rho}^2}{{\Xi}_0^2}}}})-
{\frac{{\rho}}{{\Xi}_0}}{\sqrt{1-{\frac{{\rho}^2}{{\Xi}_0^2}}}}{\Big )}$ which 
do not fulfil the required boundary condition, and since 
$Q^{\frac{3}{2}}_{\frac{1}{2}}(x){\equiv}0$ on the cut 
($Q^{-{\frac{3}{2}}}_{\frac{1}{2}}(x)$ is undefined), the next suitable solution is 
found by choosing ${\beta}=6$. This is so, since even though
$P^{-{\frac{5}{2}}}_{\frac{1}{2}}({\sqrt{1-{\frac{{\rho}^2}{{\Xi}_0^2}}}})$ does not
fulfil the required boundary condition,
$P^{\frac{5}{2}}_{\frac{1}{2}}({\sqrt{1-{\frac{{\rho}^2}{{\Xi}_0^2}}}})=
3{\sqrt{{\frac{2{\Xi}_0^5(1-{\frac{{\rho}^2}{{\Xi}_0^2}})}{{\pi}{\rho}^5}}}}$
obviously does. This solution was found in ref. [5] and corresponds to a pure 
quadrupole field since it involves the Legendre polynomial 
$P_{-3}({\cos}{\theta})=P_2({\cos}{\theta})$ (also, $Q_2({\cos}{\theta})$ is 
singular and the corresponding mode solution thus unphysical, and 
$Q_{-3}({\cos}{\theta})$ is undefined). Moreover, given the required boundary 
condition, this solution is also unique since 
$Q^{\frac{5}{2}}_{\frac{1}{2}}(x){\equiv}0$ on the cut 
($Q^{-{\frac{5}{2}}}_{\frac{1}{2}}(x)$ is undefined). 

Similarly, since the subsequent suitable solution is found by choosing
${\beta}=20$ and the $+$-sign, corresponding to a pure octopole field 
involving the Legendre polynomial $P_{-5}({\cos}{\theta})=P_4({\cos}{\theta})$, 
it would seem that all suitable solutions are given by choosing the $+$-sign 
and ${\beta}=2n(2n+1)$, $n=0,1,2,..$, corresponding to pure even multipole 
fields involving the Legendre polynomials $P_{-2n-1}({\cos}{\theta})=
P_{2n}({\cos}{\theta})$. With the required boundary condition, these solutions
are also unique, since the functions
$P^{-m-{\frac{1}{2}}}_{\frac{1}{2}}({\sqrt{1-{\frac{{\rho}^2}{{\Xi}_0^2}}}})$,
$m=0,1,2,..$, do not admit it, and since $Q^{m+{\frac{3}{2}}}_{\frac{1}{2}}(x)
{\equiv}0$ on the cut (found by using the recurrence 
relation $Q^{{\mu}+2}_{\nu}(x)+2({\mu}+1)x(1-x^2)^{-1/2}Q^{{\mu}+1}_{\nu}(x)+
({\nu}-{\mu})({\nu}+{\mu}+1)Q^{{\mu}}_{\nu}(x)=0$). Also,
$Q^{-m-{\frac{3}{2}}}_{\frac{1}{2}}(x)$ and $Q_{-m-1}({\cos}{\theta})$ are all 
undefined. Finally, since $Q_{2n}({\cos}{\theta})$ is singular for all $n$, 
the corresponding mode solutions are unphysical.

That the above extrapolation is correct, can be checked by using the formulae 
[8]
\eqa
P^{{\pm}{\sqrt{{\beta}+{\frac{1}{4}}}}}_{\frac{1}{2}}{\Big (}0{\Big )}=
2^{{\pm}{\sqrt{{\beta}+{\frac{1}{4}}}}}{\pi}^{-{\frac{1}{2}}}
{\cos}{\Big [}{\frac{\pi}{2}}{\Big (}{\frac{1}{2}}
{\pm}{\sqrt{{\beta}+{\frac{1}{4}}}}{\Big )}{\Big ]}
{\frac{{\Gamma}{\Big (}{\frac{3}{4}}{\pm}
{\frac{1}{2}}{\sqrt{{\beta}+{\frac{1}{4}}}}{\Big )}}
{{\Gamma}{\Big (}{\frac{5}{4}}{\mp}
{\frac{1}{2}}{\sqrt{{\beta}+{\frac{1}{4}}}}{\Big )}}},
\ena
\eqa
Q^{{\pm}{\sqrt{{\beta}+{\frac{1}{4}}}}}_{\frac{1}{2}}{\Big (}0{\Big )}=
-2^{{\pm}{\sqrt{{\beta}+{\frac{1}{4}}}}-1}{\pi}^{{\frac{1}{2}}}
{\sin}{\Big [}{\frac{\pi}{2}}{\Big (}{\frac{1}{2}}
{\pm}{\sqrt{{\beta}+{\frac{1}{4}}}}{\Big )}{\Big ]}
{\frac{{\Gamma}{\Big (}{\frac{3}{4}}{\pm}
{\frac{1}{2}}{\sqrt{{\beta}+{\frac{1}{4}}}}{\Big )}}
{{\Gamma}{\Big (}{\frac{5}{4}}{\mp}
{\frac{1}{2}}{\sqrt{{\beta}+{\frac{1}{4}}}}{\Big )}}}.
\ena
Since the reciprocal gamma function $1/{\Gamma}(x)$ possesses simple zeros
at $x=0,-1,-2,..$, [8], these expressions vanish for the chosen values 
${\beta}=2n(2n+1)$, $n=1,2,3,..$, and using the $+$-sign (the special case 
${\beta}=0$ was addressed previously). Moreover, no other values of ${\beta}$
will do (since the other possible values ${\beta}=2n(2n-1)$ yield functions of
the form $Q^{2n-{\frac{1}{2}}}_{\frac{1}{2}}(x){\equiv}0$), so we have really found 
all relevant solutions satisfying the boundary condition 
${\bar F}_{{\beta}{\pm}}({\Xi}_0)=0$ (using the $-$-sign rather than the $+$-sign 
yields no further zeros).

So, to summarize, said boundary condition is fulfilled by choosing the 
functions 
$P^{2n+{\frac{1}{2}}}_{\frac{1}{2}}({\sqrt{1-{\frac{{\rho}^2}{{\Xi}_0^2}}}})$ 
in equation (A.3); all other choices fail (or are irrelevant).
\end{document}